\documentclass[conference]{IEEEtran}

\usepackage{cite}
\usepackage{amsmath,amssymb,amsfonts}
\usepackage{algorithmic}
\usepackage{graphicx}
\usepackage[table,dvipsnames]{xcolor}
\usepackage[final]{microtype}
\usepackage[T1]{fontenc}
\usepackage{textcomp}

\usepackage{booktabs}
\usepackage{tabularx}
\usepackage{array}
\usepackage{makecell} 
\usepackage{placeins}
\usepackage{eso-pic}
\newcolumntype{Y}{>{\raggedright\arraybackslash}X}
\newcommand{\cmark}{\textcolor{green!50!black}{\checkmark}}
\newcommand{\xmark}{\textcolor{red!70!black}{\(\times\)}}

\usepackage[hidelinks]{hyperref}
\usepackage{url}

\hypersetup{
  pdftitle={Characterization of Request and Token Energy Costs for LLM Inference Workloads on GPU Platforms},
  pdfauthor={Prabhu Vellaisamy, Vanessa Lam, Shawn Blanton, and John Paul Shen},
  pdfkeywords={LLM inference, GPU energy efficiency, request energy, token energy, reasoning efficiency}
}
\begin{document}

\AddToShipoutPictureFG*{%
  \AtPageLowerLeft{%
    \raisebox{0.20in}[0pt][0pt]{%
      \makebox[\paperwidth][c]{%
        \parbox{0.94\paperwidth}{%
          \centering\scriptsize
          \copyright~2026 IEEE. Personal use of this material is permitted.
          Permission from IEEE must be obtained for all other uses, in any
          current or future media, including reprinting/republishing this
          material for advertising or promotional purposes, creating new
          collective works, for resale or redistribution to servers or lists,
          or reuse of any copyrighted component of this work in other works.
        }%
      }%
    }%
  }%
}

\bstctlcite{BSTcontrol}

\title{Characterization of Request and Token Energy Costs for LLM Inference Workloads on GPU Platforms}




\author{\IEEEauthorblockN{Prabhu Vellaisamy, Vanessa Lam, Shawn Blanton, and John Paul Shen}

\IEEEauthorblockA{Electrical and Computer Engineering Department\\Carnegie Mellon University\\
Pittsburgh, PA, USA}}


\maketitle
\thispagestyle{empty}
\pagestyle{empty}


\begin{abstract}

Large language model (LLM) inference serving is priced by tokens, but GPU energy is consumed over inference windows. This accounting mismatch makes token-normalized metrics incomplete, since average output-token energy can decrease even when total request energy increases. We characterize this behavior with a decomposed energy model: a fixed one-time prefill with a fixed generation setup cost, while each output-token generation step adds marginal step energy. We evaluate this LLM inference energy model on NVIDIA H100 and H200 GPUs across dense and mixture-of-experts (MoE) models, reporting both request energy and token energy as functions of model type (M), phase (P), batch size (B), context length (C), and output length (N). For Llama-3.2-1B on H200 at batch-16 and context-4K, increasing output length from 10 to 512 tokens reduces token energy from 7.46 to 0.72 J/token while total batched inference-window energy increases from 1.19 to 5.93 kJ. Batching also reduces token energy, but the gain is context-bounded: at 10 output tokens, the batch-16 to batch-1 gain falls from 6.31$\times$ at context-512 to 1.17$\times$ at context-4K. MoE models amplify this effect: sparse routing and fragmented expert execution increase fixed energy at low concurrency, while batching spreads that energy across more generated tokens and substantially narrows the dense-vs.-MoE token-energy gap. These results show that energy-aware serving should jointly optimize both request energy and token energy, rather than only reducing per-token energy cost.

\end{abstract}


\section{Introduction}

Large language model (LLM) inference is priced as a token-accounted service. Commercial APIs charge for input and output tokens~\cite{openai_pricing}, while serving systems report time-to-first-token (TTFT), time-per-output-token (TPOT), throughput, and ``goodput'' over token streams~\cite{mlperf,kwon2023pagedattention,distserve,sarathi}. Production schedulers also make decisions based on token-visible quantities, such as batch size, prompt length, and generated output length. However, GPU energy is consumed during inference windows in which the GPU is allocated, kernels are launched, memory is accessed, synchronization occurs, and the CPU orchestrates the host/runtime to keep the GPU device actively utilized~\cite{executionidle2026,character,taxbreak_todo}. Therefore, there is an accounting mismatch: the service-visible unit is the token generated, but the hardware-visible unit for energy is the inference window. A scalar unit like Joules-per-token (J/token) is useful but paints an incomplete picture, since it does not reveal whether an efficiency gain comes from reducing total energy, or from spreading a fixed inference-window energy cost across more generated tokens. For example, a longer output generation sequence can improve J/token while consuming more total Joules.

Recent industry-based economic model describes AI "tokenomics" through token utility, demand, supply, and monetization~\cite{nvidia_tokenomics_guide}. It identifies throughput per MW  (TPS/MW) and cost per token as key supply-side metrics, which further motivates the understanding of what token-normalized metrics conceal. A token is an \textit{economic accounting unit}, whereas GPU energy is incurred over a request execution window. Hence, an equal number of tokens need not impose the same physical costs, and a lower average cost per token need not indicate a lower total request cost. We ask: \textit{When does a lower token-normalized energy cost represent a true reduction in GPU energy, and when does it reflect amortization of request-window energy?}

We describe each evaluated operating point using \((M,P,B,C,N)\), where \(M\) is the model and \(P\) denotes the inference phase: TTFT-oriented for the prefill-dominated window, or decode-oriented for a long-output window. Batch size (\(B\)) specifies how many requests share an inference window~\cite{kwon2023pagedattention}, context length (\(C\)) determines prompt-processing work and KV-cache state, and output length (\(N\)) determines how many autoregressive decode steps are executed. These parameters are already visible to serving systems, but a single token-normalized metric does not capture their energy impact.

\begin{figure*}
    \centering
    \includegraphics[width=0.85\textwidth, height=8.2cm]{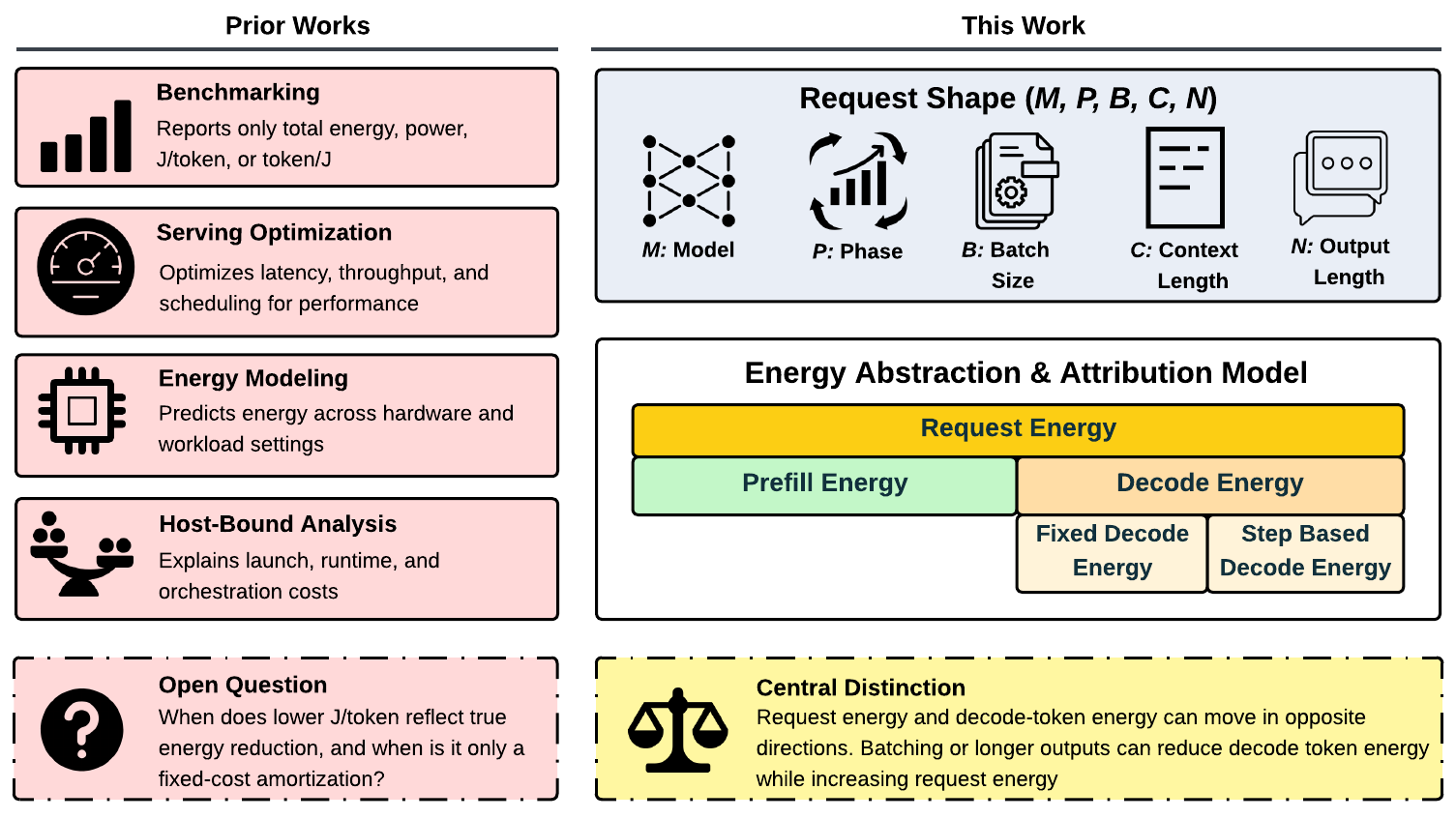}
    \caption{Request-shape energy abstraction for LLM inference. Prior work reports aggregate energy, power, J/token, tokens/J, serving performance, energy models, or host-side overheads. This work focuses on when lower J/token reflects true energy reduction versus spreading fixed energy across more tokens. We characterize each request shape with five parameters, \((M, P, B, C, N)\), denoting model, phase,  batch size, context length, and output length. Request energy decomposes into prefill and decode energy. Decode energy contains fixed energy paid once for the decode window and step energy added by each decode step. Our proposed energy abstraction explains why batching or longer outputs can reduce token energy while increasing total request energy. Both token energy and request energy must be considered in overall system optimizations.}
    \label{fig:request_shape_overall_abstraction}
\end{figure*}

\begin{figure}
    \centering
    \includegraphics[width=\linewidth]{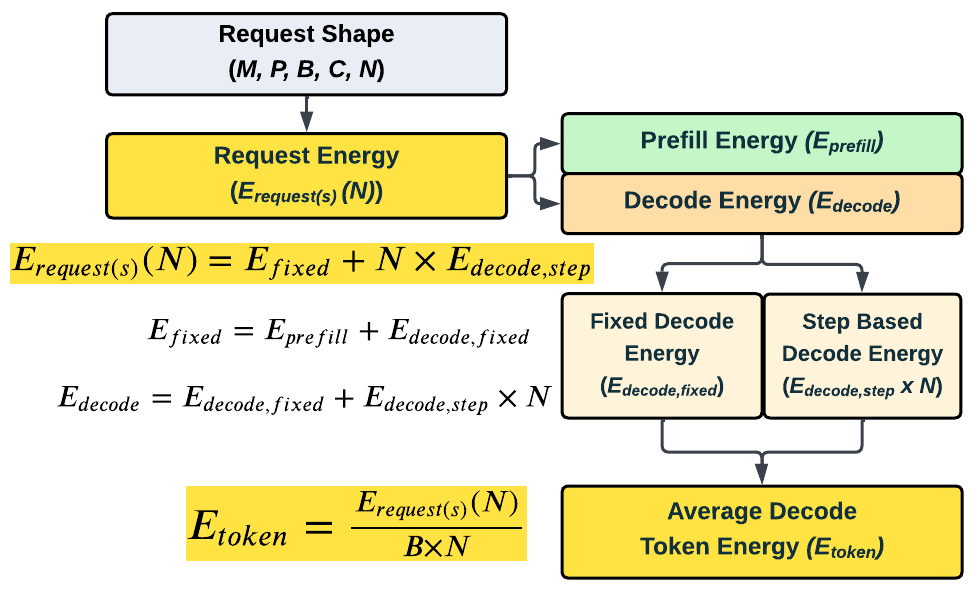}
    \caption{Energy abstraction and attribution model: request shape-based energy decomposition by separating absolute request energy from token-normalized request energy.}
    \label{fig:request_shape_energy_model}
\end{figure}

Our energy model decomposes the total energy for each inference request into two components based on the two servicing phases: \emph{prefill} phase processes the prompt, and \emph{decode} phase generates output tokens~\cite{distserve,sarathi}. 
The central abstraction in this paper is a decomposed energy model for inference request energy (see Fig. \ref{fig:request_shape_overall_abstraction} and Fig. \ref{fig:request_shape_energy_model}):
\begin{equation}
\label{eq:fixed-step}
E_{\mathrm{request(s)}}(N) = E_{\mathrm{fixed}}+ N \times\,E_{\mathrm{decode,step}}
\end{equation}
where \(E_{\mathrm{request(s)}}(N)\) is the measured GPU energy of the complete inference window containing \(B\) requests, each generating \(N\) output tokens. \(E_{\mathrm{fixed}}\) is the request-level energy independent of output length, and \(E_{\mathrm{decode,step}}\) is the marginal energy added per output token generation step for the complete batch. Conceptually, 
\begin{equation}
E_{\mathrm{fixed}}
=
E_{\mathrm{prefill}}
+
E_{\mathrm{decode,fixed}}
\end{equation}
where \(E_{\mathrm{prefill}}\) is the one-time energy used to process the prompt and construct the initial KV cache, and \(E_{\mathrm{decode,fixed}}\) represents generation-window setup and other costs that do not grow with output length.

Host/runtime orchestration, synchronization, model residency, and kernel-launch activity can contribute to both the fixed and step terms~\cite{character,taxbreak_todo}. One-time setup activity contributes to \(E_{\mathrm{fixed}}\), whereas residency energy and orchestration or launch activity that grow with inference-window duration are represented through \(E_{\mathrm{decode,step}}\). This decomposition holds model, batch size, initial context length, platform, and runtime fixed while output length varies. It is used as a local affine approximation over the measured output-length range, and for very long outputs, growing KV-cache traffic can cause marginal step energy to increase.

The corresponding average per-request energy is
\begin{equation}
E_{\mathrm{req}}
=
\frac{E_{\mathrm{request(s)}}(N)}{B}
\end{equation}
This model explains why request energy and average token energy can move in opposite directions. We define average token energy as:
\begin{equation}
\label{eq:average-token-energy}
E_{\mathrm{token}} \;=\; \frac{E_{\mathrm{request(s)}}(N)}{B \times N}
\end{equation}
This normalizes total batched request energy by \(B \times N\) generated tokens and therefore includes fixed request-level energy amortized across those tokens. Increasing output length \(N\) can reduce energy per output token by amortizing fixed request costs, while still increasing total and per-request energy. Increasing batch size \(B\) can reduce token energy by sharing fixed costs across requests, although its effect on total and per-request energy depends on the request shape and platform. Energy-aware serving must therefore track both \textit{request energy} (\(E_{\mathrm{req}}\)) and \textit{token energy} (\(E_{\mathrm{token}}\)).
Modern serving systems already manipulate request shape for latency and throughput~\cite{kwon2023pagedattention, distserve, sarathi}. We study the same surface as an \textit{energy-control} surface, asking when a shape reduces token energy by spreading fixed energy across more tokens and when this can lead to increasing the total request energy.

Prior LLM energy studies establish that inference energy depends on model, platform, runtime, and workload configuration, and commonly report average power, total energy, J/token, or tokens/J~\cite{chen2024empirical,fernandez2025energy,tokenpowerbench,samsi2023words,wilhelm2025energytoken,insights_llm_energy,energy_to_token}. Energy modeling and serving-optimization studies further show that input length, output length, hardware choice, runtime configuration, and service level objective (SLO) constraints shift the energy operating point~\cite{wilkins2024offline,sweetspots2026,dynamollm,liu2025greenllmsloawaredynamicfrequency,kakolyris2025sloawaregpufrequencyscaling,ifath2026characterizingperformanceenergytradeoffslarge}. Host-side characterization studies show that framework dispatch, CUDA-library overhead, and kernel launch costs can dominate short decode regimes~\cite{character,taxbreak_todo}, while execution-idle studies show that allocated GPUs can remain in elevated-power states even when visible activity is low~\cite{executionidle2026}. These results motivate careful energy measurement, but do not separate request energy from token energy. 
We measure LLM inference energy using the GPU cumulative NVML energy counter \cite{nvml2026}, which gives total energy over the inference window. We evaluate dense and mixture-of-experts (MoE) models on NVIDIA H100 and H200 GPUs across eager and FlashAttention runtimes~\cite{dao2022flashattention}, TTFT-oriented and long-output request regimes, batch sizes \(B\), context lengths \(C\), and output lengths \(N\) up to 8K tokens. 
This paper makes the following contributions:
\begin{itemize}
    \item \textbf{A fixed energy and step energy model.}
 We formulate measured request energy as fixed request-level energy plus marginal step energy. The fixed term includes one-time prefill and fixed generation setup, while each output-token generation step contributes marginal energy. This model explains why token energy can decrease even when request energy increases (see Fig. \ref{fig:request_shape_overall_abstraction} and Fig. \ref{fig:request_shape_energy_model}).

    \item \textbf{A request-shape characterization of LLM inference energy.}
    We measure how analysis regime, batch size, context length, and output length affect request and token energy. Across platforms, H200 consistently lowers dense-model token energy, while MoE differences depend on model and request shape.

    \item \textbf{Evidence that output length and batching spread fixed energy.}
    We show that longer outputs and larger batches can reduce average token energy (\(E_\mathrm{token}\)), but gains shrink with context length, and very long outputs can enter a regime where step energy dominates. We also show that MoE models are especially sensitive to batching, consistent with substantial routing, dispatch, and fragmented expert-execution overheads being amortized across request shape.

    \item \textbf{Attribution of floor- and kernel-active energy.}
    We combine NVML request-energy measurements with resident-floor and isolated-kernel replay measurements to explain why short, low-concurrency inference windows can be dominated by floor energy and orchestration rather than active kernel work.

    \item \textbf{Free batching for Perf/W}.
    We report prompt tokens/J to identify the free-batching regime, where additional batched prompt work fits into an already-open TTFT-oriented window and improves energy efficiency, and the point where larger contexts move the request into a compute-heavy regime. This exposes transitions that are not visible from \(T_{N=1}\), the complete \(N=1\) window time, alone.

    \item \textbf{Utility-aware reasoning-budget characterization.}
    We evaluate Qwen3-8B, DeepSeek-R1-Distill-Qwen-7B, and DeepSeek-R1-Distill-Llama-8B on MATH-500 and ARC-Challenge under 64-, 256-, and 1024-token reasoning caps to show that utility/J-maximizing reasoning budget is model- and workload-dependent, and need not coincide with accuracy-maximizing budget.

\end{itemize}

\section{Background and Related Works}

\subsection{LLM Inference Execution}


\begin{table*}[t]
\caption{Comparison of this work against prior LLM inference energy studies.}
\label{tab:related_energy}
\centering
\footnotesize
\setlength{\tabcolsep}{3.0pt}
\renewcommand{\arraystretch}{1.2}

\begin{tabularx}{\dimexpr\textwidth-2pt\relax}{@{} >{\raggedright\arraybackslash}p{2.7cm} X *{8}{c} @{}}
\toprule
\textbf{Work / Group} &
\textbf{Focus} &
\textbf{Phase} &
\textbf{Batch} &
\textbf{Ctx.} &
\textbf{Out.} &
\textbf{HW} &
\textbf{MoE} &
\textbf{Req./Token} &
\textbf{Window} \\
\midrule

Energy benchmarks~\cite{samsi2023words,chen2024empirical,tokenpowerbench} &
Power/energy measurement and token-normalized inference metrics. &
\cmark & \cmark & \cmark & \cmark & \cmark & \cmark & \xmark & \xmark \\

Serving and optimization~\cite{fernandez2025energy,dynamollm} &
Energy impact of frameworks, batching, decoding, hardware, and system control. &
\cmark & \cmark & \cmark & \cmark & \cmark & \cmark & \xmark & \xmark \\

Input/output models~\cite{wilkins2024offline,wilhelm2025energytoken,sweetspots2026} &
Energy models and operating curves over input/output token counts. &
\cmark & \xmark & \cmark & \cmark & \cmark & \cmark & \xmark & \xmark \\

Host-side overhead~\cite{character,taxbreak_todo,executionidle2026} &
Kernel-launch and dispatch decomposition, execution-idle high-power states. &
\cmark & \cmark & \cmark & \xmark & \cmark & \xmark & \xmark & \cmark \\

\midrule

\textbf{This Work} &
\textbf{Abstraction of LLM inference energy costs by differentiating inference request energy and average token energy, and shows how schedulers should account for both for efficient inference service.
} &
\textbf{\cmark} & \textbf{\cmark} & \textbf{\cmark} & \textbf{\cmark} &
\textbf{\cmark} & \textbf{\cmark} & \textbf{\cmark} & \textbf{\cmark} \\

\bottomrule
\end{tabularx}

\vspace{4pt}
\begin{flushleft}
\footnotesize
\textbf{Legend:} Phase = prefill/decode; Batch = batch size; Ctx. = context length; Out. = output length; HW = hardware/platform comparison; req./token = explicit separation of total request energy from token-normalized energy; Window = inference-window or decode-window amortization mechanism.
\end{flushleft}
\end{table*}

LLM inference has two phases: \emph{prefill} and \emph{decode}. In prefill, the model processes the full input prompt in one forward pass and populates the key-value (KV) cache. As prompt tokens are processed together, prefill exposes substantial token-level parallelism and is typically dominated by large batched GEMMs~\cite{taxbreak_todo}. In decode, the model generates one output token per forward pass. Decode is sequential across output tokens because each step depends on the KV state produced by previous steps. Its cost, therefore, grows with output length (\(N\)) and with the accumulated KV state, which scales with batch size (\(B\)) and context length (\(C\)).

Modern serving systems exploit this phase asymmetry through scheduling and memory management. PagedAttention virtualizes the KV cache into fixed-size blocks and enables continuous batching~\cite{kwon2023pagedattention}. DistServe separates prefill and decode workers to reduce phase interference~\cite{distserve}, while Sarathi-Serve chunks long prefills and schedules them alongside decode work~\cite{sarathi}. These systems primarily optimize latency and throughput, but their packing decisions also change inference request energy.

\subsection{Related Works}

\subsubsection{LLM inference energy benchmarking}
Prior work has established GPU energy as a first-class metric for LLM inference. Words-to-Watts~\cite{samsi2023words} measures inference energy across language tasks and serving stacks, while Chen~et~al.~\cite{chen2024empirical} report power and energy variation across GPU generations under matched workloads. TokenPowerBench~\cite{tokenpowerbench} extends this direction with phase-aligned prefill/decode measurements across batch size, context length, inference engine, and parallelism. Wilhelm~et~al.~\cite{wilhelm2025energytoken} advocate energy-per-token as a first-class metric beyond accuracy, and Fernandez~et~al.~\cite{fernandez2025energy} quantify how framework choice, batching, decoding strategy, quantization, hardware platform, and online/offline serving mode shift inference energy. These studies provide strong measurement foundations, but primarily report aggregate or token-normalized quantities such as average power, total joules, J/token, or tokens/J. Such metrics do not distinguish reductions in step energy from spreading fixed energy across more tokens.

\subsubsection{Token economics}
NVIDIA’s tokenomics framework~\cite{nvidia_tokenomics_guide} links token utility and demand to infrastructure supply and monetization. It emphasizes model-system-software co-design and metrics such as cost per token and throughput per megawatt. This work is complementary: rather than modeling price or total cost of ownership, we isolate GPU energy and decompose the complete inference window into fixed and output-length-dependent components. This separation reveals that a favorable token-normalized metric is achieved by amortization rather than by a lower absolute energy.

\subsubsection{Energy modeling and serving optimization}
A complementary line of work models LLM energy as a function of workload shape and serving configuration. Wilkins~et~al.~\cite{wilkins2024offline} fit per-LLM energy and runtime models for offline energy-optimal scheduling on heterogeneous CPU-GPU systems. SweetSpot~\cite{sweetspots2026} derives an analytical efficiency model from Transformer compute and memory complexity and identifies non-linear input/output-length regimes. Cluster-scale work such as DynamoLLM~\cite{dynamollm} designs heterogeneous-instance and frequency-scaling policies under SLO constraints. These works show that energy efficiency depends strongly on workload shape, system configuration, and deployment policy, but they do not focus on cases where token-level efficiency improves while total request energy increases.

\subsubsection{Host overhead and execution-idle}
Prior CPU-GPU coupling work identifies host orchestration and kernel-launch overheads using kernel launch and queue time~\cite{character}, while TaxBreak~\cite{taxbreak_todo} decomposes inference overhead into Python translation, ATen dispatch, CUDA-library overhead, and kernel-launch components. Execution-idle studies further show that allocated GPUs can remain in elevated-power states even when visible kernel activity is low~\cite{executionidle2026}. These results show that an inference window can carry fixed orchestration energy. However, they primarily expose host-bound behavior through latency and overhead decomposition, not through request-level energy or performance per watt.


\subsection{This Work}

This work makes the distinction between \emph{request energy} and \emph{token energy} the main object of characterization (Table \ref{tab:related_energy}). We focus on request shapes where batching or longer outputs reduce token energy by spreading fixed energy across more tokens, but can still increase total request energy by extending the decode window. We capture this with a fixed energy and step energy model (Eq. \ref{eq:fixed-step}) and use floor, kernel-active, and residual attribution to explain selected trends.

\section{Experimental Methodology}
\label{sec:methodology}

Our methodology measures how request shape impacts GPU energy and token-normalized request efficiency. For each configuration, we vary the request-shape parameters and measure GPU energy over the corresponding inference window using the cumulative NVML energy counter.

\subsection{Request Shape and Energy Metrics}
\label{sec:metrics}

We denote a measured configuration by \((M,P,B,C,N)\), with NVML energy measuring over a complete inference request. We define request energy as the total GPU energy consumed by each individual inference request:
\begin{equation}
E_{\mathrm{req}}
=
\frac{E_{\mathrm{request(s)}}(N)}{B}
\end{equation}
\(E_{\mathrm{request(s)}}(N)\) is the measured energy of the complete batched request window. Its fixed term includes one-time prefill and fixed generation setup, while its output-length-dependent term captures marginal generation-step energy.

We focus on average token energy \(E_{\mathrm{token}}\) as defined in Eq.~\ref{eq:average-token-energy}, which normalizes total batched request energy by generated output tokens, \(E_{\mathrm{request(s)}}(N)
=E_{\mathrm{fixed}}+ N \times E_{\mathrm{decode,step}}\), which also includes prefill (Fig. \ref{fig:request_shape_energy_model}).

\subsection{NVML Energy Measurement}
\label{sec:nvml_measurement}

Dynamic power and energy telemetry on NVIDIA GPUs is exposed through the NVIDIA Management Library (\texttt{NVML})~\cite{nvml2026}. We use two \texttt{NVML} interfaces. \texttt{nvmlDeviceGetTotalEnergyConsumption} reports cumulative energy consumption in mJ. Differences between synchronized readings at inference-window boundaries provide GPU energy and its associated circuitry during the measured window (\(E_{\mathrm{NVML}}\)). \texttt{nvmlDeviceGetPowerUsage} reports power in mW averaged over a one-second interval. The cumulative energy-counter difference is used as the primary energy metric since the power samples averaging one-second intervals cannot temporally resolve short decode windows, small batches, or transitions between GPU execution and host/runtime gaps~\cite{mcdaniel2026finegrainedpowerenergyattribution, tran2026wattchmen}. Hence, power readings are polled at a nominal 50\,ms cadence as smoothed diagnostics through \texttt{pynvml}~\cite{nvidia-ml-py}. Calling \texttt{NVML} directly from the profiling process also avoids the startup and parsing overheads of repeated \texttt{nvidia-smi} queries.

\subsection{Measurement Protocol}
\label{sec:execution_protocol}

Each configuration performs warmup iterations followed by profiled iterations using \texttt{torch.profiler}~\cite{paszke2019pytorch, torch_profiler2026}. By default, we use \(W=10\) warmup runs and \(R=10\) profiled runs. Mean latency is reported over the profiled runs, and cumulative‑counter energy is measured over the last five profiled runs and normalized per inference to maintain tractable measurement runtimes. 

To ensure deterministic request shapes, prompts are synthesized as random token tensors drawn uniformly from each model's vocabulary while excluding EOS IDs. The same input tensor is reused across warmup and measured iterations. Attention masks are all ones, so the requested context length \(C\) is the actual attended sequence length. Further, \(C\) is clamped to the model's configured maximum, and EOS termination is disabled so that each request generates exactly \(N\) output tokens.

We use \(N=1\) as a TTFT-oriented, prefill-dominated configuration and \(N>1\) to measure complete fixed-length inference windows. The \(N=1\) measurement includes prompt processing, generation setup, and first-token production, while each \(N>1\) measurement includes the same one-time work followed by additional decode steps. Consequently, all request- and token-energy results are computed from the measured energy of the complete inference window. Floor and per-kernel replay measurements are used for diagnostic attribution and are described next.

\subsection{Diagnostic Floor and Kernel Attribution}
\label{sec:diagnostic_attribution}

We use lightweight attribution to explain measured trends. First, we measure resident floor power from loading the model into GPU memory, initializing the runtime, synchronizing CUDA, and issuing no kernels. We sample GPU package power through \texttt{pynvml} at a nominal 50\,ms cadence. When the cumulative NVML energy counter is available, we also take a longer idle measurement window counter difference as a cross-check and use the energy-counter-derived power as the primary resident-floor value. We use the measured resident floor energy as:
\begin{equation}
    E_{\mathrm{floor}} = P_{\mathrm{floor}} \cdot T_{\mathrm{window}}
\end{equation}

This helps explain why short, low-concurrency decode windows can have high \(E_{\mathrm{token}}\). Second, we estimate kernel-active energy using isolated replay. Following the replay-based attribution strategy used in TaxBreak~\cite{taxbreak_todo}, we keep a database of unique kernels recorded in each model run along with their metadata (name, launch shape, input shape, and dtype). Each unique kernel signature is replayed in isolation in a child process. The replay first runs a 500\,ms warmup, then runs three 500\,ms measurement windows whose boundaries are emitted by the child process and observed by the parent. The parent reads \texttt{nvmlDeviceGetTotalEnergyConsumption} at each boundary and also records \texttt{nvmlDeviceGetPowerUsage} as a smoothed diagnostic. Because Hopper power readings are averaged over one second, they are not used as an independent consensus channel for the 500\,ms replay windows. Instead, replay reliability is evaluated from the cumulative energy-counter windows: at least two effective windows must remain, and the inter-window variation of the energy-derived net power must be at most 5\%. For a kernel \(k\), the replay-derived active energy contribution is:
\begin{equation}
E_{\mathrm{active},k} = P_{\mathrm{net},k} t_{k} \nu_{k}
\label{eq:active_energy}
\end{equation}
where \(P_{\mathrm{net},k}\) is the energy-counter-derived active power above the local floor, \(t_k\) is the mean production duration, and \(\nu_k\) is the production dispatch count. We define \(E_{\mathrm{active}}=\sum_k E_{\mathrm{active},k}\).
For these diagnostic cases, we report:
\begin{equation}
    E_{\mathrm{NVML}} = E_{\mathrm{floor}} + E_{\mathrm{active}} + E_{\mathrm{residual}}
\end{equation}

And hence, \(E_{\mathrm{residual}}=E_{\mathrm{NVML}}-E_{\mathrm{floor}}-E_{\mathrm{active}}\). Because floor and isolated-replay estimates are not strictly additive under production overlap, the diagnostic residual may be signed. The residual is not redistributed to kernels. Because isolated replay cannot perfectly reproduce the production inference window, this attribution is used only to support interpretation of how fixed energy is spread across generated tokens.


\section{Experimental Setup}
\label{sec:experiment}

\subsection{Hardware Systems}

We evaluate the workloads on two single-GPU NVIDIA Hopper-class platforms from a shared
research cluster:

\begin{itemize}
    \item \textbf{H100 platform:} Intel Xeon 8480C, 56 cores
    @ 2.0/3.8\,GHz, PCIe Gen5, paired with an NVIDIA H100 SXM
    GPU with 80\,GB HBM3 and 700\,W nominal TDP.
    \item \textbf{H200 platform:} Intel Xeon Gold 6538Y+, 32 cores,
    paired with an NVIDIA H200 NVL GPU with 141\,GB HBM3e and
    600\,W nominal TDP.
\end{itemize}
Table~\ref{tab:hopper_platforms} summarizes the GPU-level differences relevant to LLM inference. H100 SXM and H200 NVL use the same Hopper generation, but they are not matched devices: H200 NVL has a larger and higher-bandwidth HBM3e subsystem, while the H100 SXM has a higher advertised peak Tensor-compute rate and a higher configurable TDP.
\begin{table}[t]
\centering
\caption{Differences between the evaluated Hopper GPU platforms. Peak Tensor throughput assumes sparsity. Product specifications are from NVIDIA~\cite{nvidiah100specs,nvidiah200specs}. Prior work shows that energy efficiency depends on both GPU architecture and AI workload under power limits~\cite{mayrpoweraware2026}, while memory bandwidth and DVFS can materially affect that relationship~\cite{ujeniyah200power2026}.}
\label{tab:hopper_platforms}
\footnotesize
\setlength{\tabcolsep}{3.5pt}
\renewcommand{\arraystretch}{1.12}
\begin{tabular}{@{}lcc@{}}
\toprule
\textbf{Feature} & \textbf{H100 SXM} & \textbf{H200 NVL} \\
\midrule
Compute generation & Hopper (GH100) & Hopper (GH100) \\
Peak BF16/FP16 Tensor & 1,979 TFLOPS & 1,671 TFLOPS \\
GPU memory & 80\,GB HBM3 & 141\,GB HBM3e \\
Memory bandwidth & 3.35\,TB/s & 4.8\,TB/s \\
Maximum configurable TDP & 700\,W & 600\,W \\
Form factor & SXM & PCIe, dual-slot air-cooled \\
\bottomrule
\end{tabular}
\end{table}
All runs use one GPU and six CPU cores per GPU. Host-memory reservations vary from 32 to 192 GB to accommodate model- and trace-dependent profiler memory requirements.

We selected H100 and H200 as commercially available Hopper-generation platforms that are representative of current large-model deployments. The H200 provides greater memory capacity and bandwidth, features that affect decode energy by changing the cost of KV-cache traffic and long-context inference. This is a system-level comparison rather than an isolated memory experiment, since the two test platforms also differ in host CPU, GPU form factor, and nominal TDP. We consequently attribute cross-platform energy differences to the complete platform configuration, and relate the observed trends to these features in the H100 vs. H200 results discussion.

\subsection{System Software}

We use a Python 3.13 environment with PyTorch \texttt{2.9.0+cu128}, \texttt{cuda12.6/toolkit} module, and Transformers 4.57.3. Unless stated otherwise, runs use BFloat16 eager execution. FlashAttention-2 is enabled only for the corresponding sweep. Request energy is measured using NVML through the \texttt{pynvml} Python interface, while power telemetry is retained only as a diagnostic. Trace collection uses \texttt{torch.profiler}. Per kernel replay runs in a child process under \texttt{nsys profile --trace=cuda,nvtx} to validate CUDA dispatch and collect kernel activity. 

\subsection{LLM Workloads}

We evaluate dense and mixture-of-experts LLMs in BFloat16. Dense LLMs include Llama-3.2-1B, Llama-3.2-3B, and Llama-3.1-8B~\cite{grattafiori2024llama}. MoE models include OLMoE‑1B‑7B and Qwen1.5‑MoE‑A2.7B, which differ in expert organization. OLMoE uses 64 routed experts per layer with top‑8 routing and no shared experts, whereas Qwen1.5‑MoE uses 64 experts per layer with 4 shared and 60 routed experts under top‑4 routing. We use Hugging Face’s default per‑expert independent matrix multiplications and disable token dropping. Prompts are synthesized to the target context length to eliminate tokenizer‑induced length variation.

\section{Experimental Results and Observations}

First, we vary output length and batch size to illustrate how decode‑window costs amortize and where that amortization ceases to hold. (Section~\ref{sec:5.1}). We then compare H100 and H200 to separate platform scaling from request-shape structure (Section~\ref{sec:5.2}), contrast prefill and decode as distinct operating regimes (Section~\ref{sec:5.3}), and use per-kernel attribution to explain floor-dominated cases (Section~\ref{sec:5.4}). We close by summarizing the scheduler-facing implications (Section~\ref{sec:5.5}). Section~\ref{app:utility} then extends the analysis to utility-aware reasoning, examining how capped reasoning budgets trade task accuracy against request energy and utility per joule across models, benchmarks, and GPU platforms.

\subsection{Output Length Is a First-Class Energy Axis}
\label{sec:5.1}

\begin{figure}[t]
    \centering
    \includegraphics[width=\linewidth, height=5.5cm]{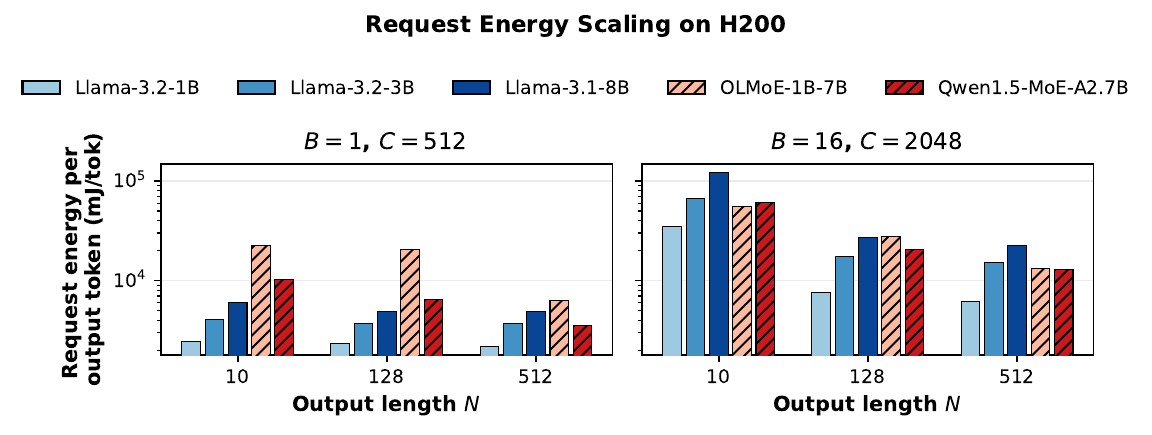}
    \caption{Request energy scaling (mJ/token) across varying output generation lengths. The evaluation contrasts a low-load scenario (left) against a high-load scenario (right) on a single NVIDIA H200 GPU. The y-axis is log-scaled.}
    \label{fig:energy_scaling}
\end{figure}

\begin{figure}[t]
    \centering
    \includegraphics[width=\linewidth, height=5.5cm]{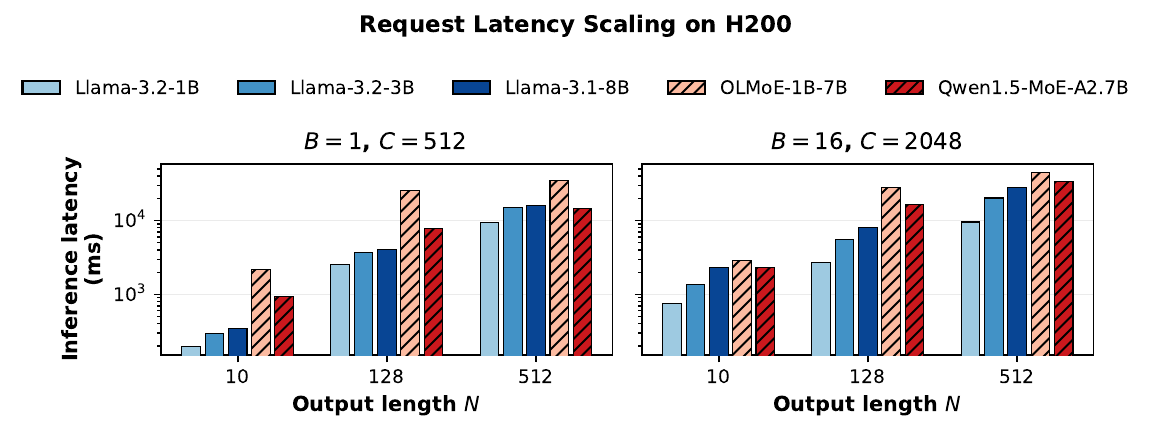}
    \caption{Inference latency scaling (ms) across varying output generation lengths. The evaluation contrasts a low-load scenario (left) against a high-load scenario (right) on a single NVIDIA H200 GPU. The y-axis is log scaled.}
    \label{fig:latency_scaling}
\end{figure}

Fig.~\ref{fig:energy_scaling} plots token energy across output lengths \(N \in \{10, 128, 512\}\) for two H200 operating points: low load (\(B{=}1\), \(C{=}512\)) and high load (\(B{=}16\), \(C{=}2048\)). Fig.~\ref{fig:latency_scaling} shows the corresponding latency results. Together, they show that output length plays a crucial role both as a performance axis and an energy-accounting axis.

\subsubsection{Low-load amortization is modest for dense models}

At (\(B{=}1\), \(C{=}512\)), increasing \(N\) from 10 to 512 tokens reduces token energy by only 1.1\(\times\) for Llama-3.2-1B (2{,}431\(\to\)2{,}182\,mJ/token) and 1.2\(\times\) for Llama-3.1-8B (6{,}011\(\to\)4{,}866\,mJ/token). This near-flat response means that extending the decode window does not substantially change the energy paid per generated token in this regime. The request remains low-concurrency and underutilized, so a large amount of fixed energy is spread across relatively little token work.

In contrast, the MoE models in Fig.~\ref{fig:energy_scaling} show a stronger reduction in token energy as $N$ increases. This behavior is consistent with a larger output-length-amortizable component arising from their larger total parameter footprint and the routing, dispatch, and fragmented expert-execution overheads identified by
TaxBreak~\cite{taxbreak_todo}. These mechanisms can contribute to both the request window and the per-decode-step costs. Therefore, the observed MoE scaling indicates greater amortization of
low-concurrency overhead.

\subsubsection{High-load amortization is stronger} 

At (\(B{=}16\), \(C{=}2048\)), the same output-length sweep yields a 5.7\(\times\) reduction for Llama-3.2-1B (2{,}198\(\to\)386\,mJ/token) and a 5.4\(\times\) reduction for Llama-3.1-8B (7{,}588\(\to\)1{,}398\,mJ/token). Here, fixed energy is spread across more generated tokens, so token energy improves even though the decode window consumes more total energy.

\subsubsection{A linear request-energy fit separates fixed energy and step energy}

For each fixed \((M,B,C)\) configuration, we fit the measured complete inference-window energy at \(N \in \{10,128,512\}\) to \(E_{\mathrm{request}}(N)=E_{\mathrm{fixed}}+N E_{\mathrm{step}}\). The intercept estimates energy independent of output length, while \(E_{\mathrm{step}}\) is the incremental energy of one full-batch decode step. Table~\ref{tab:decode_linear_fit} reports \(E_{\mathrm{step}}/B\), the corresponding incremental energy per generated token.

The model captures the overall increase in request energy, but its relative error can be substantial at \(N{=}10\), where the request is shortest. Table~\ref{tab:decode_linear_fit} further reports residuals at each measured length. We use the model only as a local decomposition over the measured range, not as evidence that marginal step energy is constant outside it.

\begin{table}[t]
\centering
\caption{Representative H200 affine fits over \(N\in\{10,128,512\}\). Residuals are \((E_{\mathrm{fit}}-E_{\mathrm{measured}})/E_{\mathrm{measured}}\).}
\label{tab:decode_linear_fit}
\footnotesize
\setlength{\tabcolsep}{2.5pt}
\renewcommand{\arraystretch}{1.12}
\begin{tabular}{@{} l c c c c c c c @{}}
\toprule
\textbf{Model} &
\textbf{\(B\)} &
\textbf{\(C\)} &
\textbf{\(E_{\mathrm{fixed}}\)} &
\textbf{\(E_{\mathrm{step}}/B\)} &
\multicolumn{3}{c}{\textbf{Residual (\%)}} \\
& & &
\textbf{(J)} &
\textbf{(mJ/token)} &
\textbf{\(N{=}10\)} &
\textbf{\(N{=}128\)} &
\textbf{\(N{=}512\)} \\
\midrule
Llama-3.2-1B & 1  & 512  & 12.11  & 2163.56 & +38.79 & -4.09 & +0.26 \\
Llama-3.2-1B & 16 & 2048 & 276.78 & 351.06  & -5.34  & +2.53 & -0.18 \\
Llama-3.2-3B & 16 & 512  & 30.63  & 437.17  & -28.34 & +5.95 & -0.34 \\
Llama-3.1-8B & 16 & 512  & 83.73  & 604.84  & -27.81 & +7.38 & -0.42 \\
\bottomrule
\end{tabular}
\end{table}

\begin{table}[t]
\centering
\caption{Batching gain for complete H200 requests at \(N{=}10\), reported as \(E_{\mathrm{token}}(B{=}1)/E_{\mathrm{token}}(B{=}16)\).}
\label{tab:batch_gain_context}
\small
\setlength{\tabcolsep}{5pt}
\renewcommand{\arraystretch}{1.15}
\begin{tabular}{@{} l c c c c @{}}
\toprule
\textbf{Model} & \textbf{\(C{=}512\)} & \textbf{\(C{=}1024\)} & \textbf{\(C{=}2048\)} & \textbf{\(C{=}4096\)} \\
\midrule
Llama-3.2-1B & 6.31\(\times\) & 3.19\(\times\) & 1.85\(\times\) & 1.17\(\times\) \\
Llama-3.2-3B & 4.67\(\times\) &  2.66\(\times\) & 1.70\(\times\) & 1.22\(\times\) \\
Llama-3.1-8B & 3.85\(\times\) & 2.15\(\times\) & 1.47\(\times\) & OOM \\
OLMoE-1B-7B & 12.59\(\times\) & 11.88\(\times\) & 6.41\(\times\) & 3.93\(\times\) \\
Qwen1.5-MoE-A2.7B & 6.60\(\times\) & 4.75\(\times\) & 3.18\(\times\) & 1.90\(\times\) \\
\bottomrule
\end{tabular}
\end{table}

\subsubsection{Batch gain is context-bounded} Batch size (\(B\)) and output length (\(N\)) can both spread fixed energy across more generated tokens, but context length increases step energy. For Llama-3.2-1B at \(N{=}512\), the \(B{=}16\) token-energy advantage over \(B{=}1\) is 11.2\(\times\) at \(C{=}512\), but shrinks to 6.0\(\times\) at \(C{=}2048\) and 3.7\(\times\) at \(C{=}4096\). Table~\ref{tab:batch_gain_context} shows the same context-bounded pattern for short-output decode at \(N{=}10\), where the absolute gains are smaller because fewer output tokens are available to spread the fixed energy. As context grows, the KV state accessed during each decode step grows as well, moving the workload from a regime dominated by fixed energy toward one dominated by step energy. This trend is consistent with growing HBM demand: each decode step reads a larger KV cache as context length increases, so the memory subsystem becomes a larger component of step energy and leaves less fixed energy for batching to spread across tokens.

\subsubsection{Very long outputs reveal a second regime} The linear fit is not intended as a global model for arbitrarily long generations. Extended request runs with \(N \in \{1024,4096,8192\}\) show that token energy is not guaranteed to keep decreasing with output length (Table~\ref{tab:long_output_decode}). Llama-3.2-3B/H200 at (\(B{=}16\), \(C{=}512\)) reaches 337\,mJ/token at \(N{=}128\), but rises to 1{,}600\,mJ/token at \(N{=}8192\). Llama-3.1-8B/H200 at the same request shape rises from 548\,mJ/token to 2{,}358\,mJ/token. Output length therefore creates two regimes: one where more generated tokens reduce token energy by spreading fixed energy, and one where step energy dominates. The second regime is consistent with the growing cache and sustained active execution of long autoregressive generation, since each additional token both extends the decode window and increases the state accessed by subsequent steps. We therefore interpret the rise as step energy overtaking the fixed component.

\begin{table}[!t]
\centering
\caption{Long-output request energy per output token on H200 at \(B{=}16\), \(C{=}512\). Values are \(E_{\mathrm{request(s)}}(N)/(B \times N)\).}
\label{tab:long_output_decode}
\small
\setlength{\tabcolsep}{14pt}
\renewcommand{\arraystretch}{1.2}
\begin{tabular}{@{} ccc @{}}
\toprule
\textbf{\(N\)} & \textbf{Llama-3.2-3B} & \textbf{Llama-3.1-8B} \\
\midrule
128  & 337\,mJ/token   & 548\,mJ/token  \\
512  & 372\,mJ/token   & 582\,mJ/token   \\
1024 & 455\,mJ/token   & 691\,mJ/token   \\
4096 & 945\,mJ/token   & 1,394\,mJ/token \\
8192 & 1,600\,mJ/token & 2,358\,mJ/token \\
\bottomrule
\end{tabular}
\end{table}

\subsubsection{Batching amortizes MoE execution overheads}
At low load (\(B{=}1\), \(C{=}512\), \(N{=}10\)), OLMoE-1B-7B costs 22{,}476\,mJ/token and Qwen1.5-MoE-A2.7B costs 10{,}277\,mJ/token. These are 9.2\(\times\) and 2.5\(\times\) higher than their active-parameter-comparable dense counterparts, Llama-3.2-1B at 2{,}431\,mJ/token and Llama-3.2-3B at 4{,}093\,mJ/token. Prior TaxBreak analysis identifies routing, dispatch, and fragmented expert execution as important MoE overheads~\cite{taxbreak_todo}. Our low-load energy gap is consistent with those mechanisms rather than active parameter count alone. The observed pattern is consistent with insufficient independent token work to fill the GPU effectively, leaving a relatively expensive under-filled execution window.

At high load (\(B{=}16\), \(C{=}2048\), \(N{=}10\)) the picture changes but does not collapse to a single ranking. OLMoE-1B-7B costs 3{,}440\,mJ/token compared to Llama-3.2-1B at 2{,}198\,mJ/token, leaving a 1.6\(\times\) active-parameter gap. Qwen1.5-MoE-A2.7B costs 3{,}781\,mJ/token and becomes 1.11\(\times\) lower in token energy than Llama-3.2-3B at 4{,}216\,mJ/token. Relative to Llama-3.1-8B, a total-parameter-comparable dense reference at 7{,}588\,mJ/token, OLMoE-1B-7B and Qwen1.5-MoE-A2.7B are 2.2\(\times\) and 2.0\(\times\) lower in token energy, respectively. The observed reduction is consistent with batching amortizing MoE routing and dispatch overheads, but the crossover depends on which dense baseline is used. Larger batches can provide more concurrent expert work and make fragmented expert kernels and dispatches less under-filled, allowing the GPU to remain active more consistently~\cite{taxbreak_todo}. The model-specific crossover indicates that routing structure and execution regularity remain relevant alongside active-parameter count.

\subsubsection{Batching consistently improves token energy in the measured space}
Across measured \((M,C,N)\) configurations on each GPU, increasing batch size from \(B{=}1\) to \(B{=}16\) reduces token energy. For OLMoE-1B-7B on H200 at
\(C{=}512\), \(N{=}10\), token energy falls from 22{,}476 to 1{,}786\,mJ/token (12.6\(\times\)). Llama-3.2-1B at the same shape falls from 2{,}431 to 386\,mJ/token (6.3\(\times\)). Under-filled batches therefore leave substantial token-energy efficiency unrealized, although latency budget and queueing policy still determine whether batching is usable.

\textbf{Key Insight.} \textit{Output length and batching reduce token energy only while fixed energy dominates. Context growth and very long outputs move the workload towards a regime where step energy dominates, so token energy and request energy must be considered together.}

\subsection{Platform Energy Gains Depend on Model and Request Shape}
\label{sec:5.2}
\begin{figure}[!t]
    \centering
    \includegraphics[width=\linewidth, height=9.8cm]{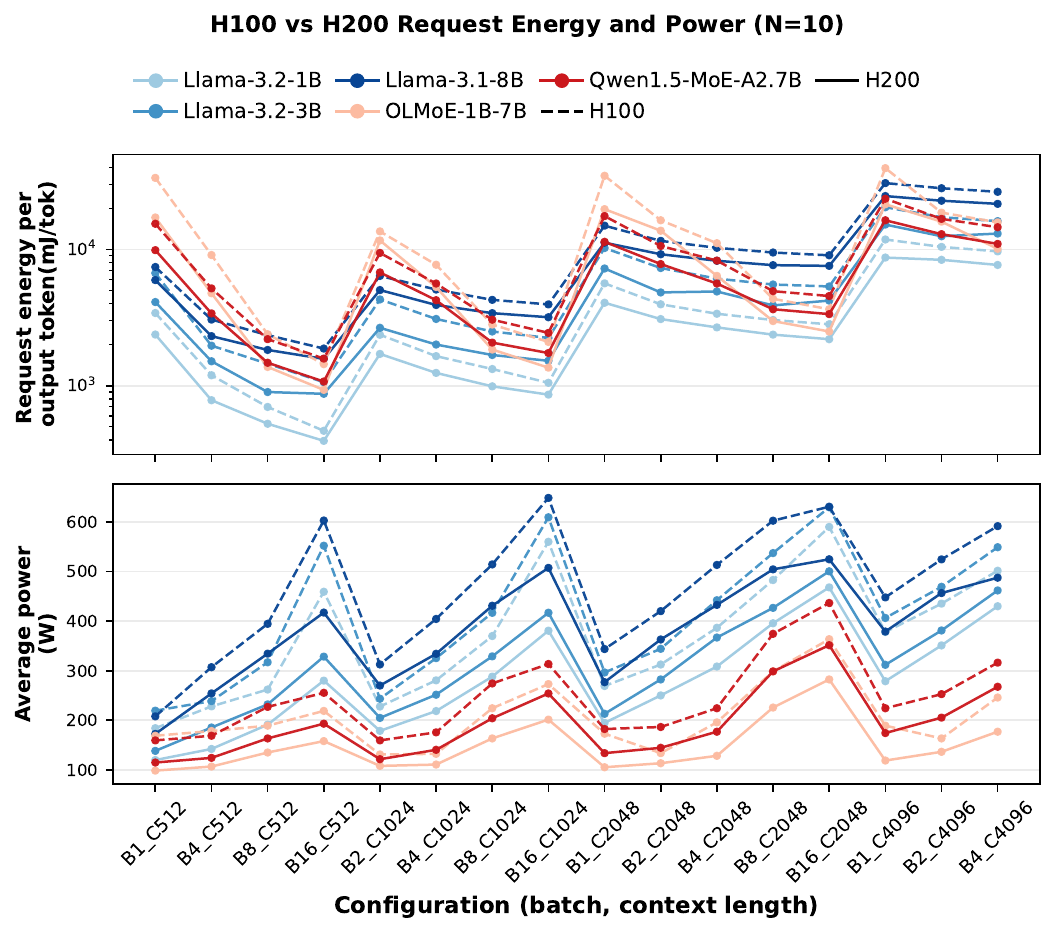}
    \caption{Request energy per output token and power scaling on H100 (dashed) vs. H200 (solid) GPUs for output length \(N=10\). This figure illustrates the scaling behavior of dense and mixture-of-experts model architectures across varying batch sizes \(B\) and context lengths \(C\).}
    \label{fig:h100_h200_comparison}
\end{figure}

Fig.~\ref{fig:h100_h200_comparison} compares H100 and H200 at \(N{=}10\) across evaluated \((B,C)\) configurations. Short-output requests make fixed request-level energy more visible, allowing us to separate platform scaling from request-shape structure.

\subsubsection{H200 lowers token energy for dense models}
Across matched Llama-3.2-1B, Llama-3.2-3B, and Llama-3.1-8B configurations, H200 reduces token energy to 0.61-0.84\(\times\) of H100. For Llama-3.2-3B at \(B{=}1\), \(C{=}512\), token energy drops from 6{,}741\,mJ/token on H100 to 4{,}093\,mJ/token on H200, a 1.65\(\times\) reduction. The smallest dense-model gain is Llama-3.1-8B at \(B{=}16\), \(C{=}2048\), where the H200/H100 ratio is 0.84\(\times\). This is a higher-load regime on both platforms, the advantage from fixed energy contributes less to the total.

The shift in sampled diagnostic power is smaller than the energy shift. Across the plotted complete \(N{=}10\) configurations, the median of the smoothed NVML power readings is 312.5\,W on H100 versus 216.4\,W on H200. Request shape still determines how long the GPU remains active, how many requests share the window, and how much KV state is accessed. The documented hardware differences make this a platform-level effect rather than a memory-only comparison. Both devices use the Hopper/GH100 generation, but H200 NVL provides 141\,GB HBM3e and 4.8\,TB/s bandwidth, compared with 80\,GB HBM3 and 3.35\,TB/s on H100 SXM. Conversely, H200 NVL has a lower advertised peak Tensor-compute rate and lower maximum configurable TDP~\cite{nvidiah100specs,nvidiah200specs}. The H200 NVL's lower configurable power envelope and different form factor are consistent with its lower sampled diagnostic power, while its higher HBM bandwidth may reduce execution time for memory-bound decode steps. Because the platforms also differ in host CPU and GPU form factor, the experiment cannot isolate the contribution of any individual feature. Prior H100/H200 studies likewise find that energy efficiency depends on workload, memory behavior, and power-management effects~\cite{mayrpoweraware2026,ujeniyah200power2026}.

\subsubsection{MoE platform gains are request-shape dependent} At (\(B{=}1\), \(C{=}2048\)), OLMoE-1B-7B drops from 27{,}671\,mJ/token on H100 to 22{,}035\,mJ/token on H200, a 0.80\(\times\) ratio. Qwen1.5-MoE-A2.7B drops from 18{,}283 to 12{,}007\,mJ/token, a 0.66\(\times\) ratio. The direction is not universal: at \((B{=}16\), \(C{=}512\)), OLMoE-1B-7B is 1{,}786\,mJ/token on H200 versus 1{,}462\,mJ/token on H100, a 1.22\(\times\) ratio, despite lower sampled power on H200 (123\,W versus 221\,W). At (\(B{=}16\), \(C{=}2048\)), its token energy is nearly equal across platforms, 3{,}440\,mJ/token on H200 vs. 3{,}521\,mJ/token on H100.

\subsubsection{The request-shape trend is qualitatively preserved across the two measured platforms} Although H200 generally shifts dense-model configurations downward in absolute energy, it does not alter the main structure of the energy surface. Low-batch and low-output decode remains expensive per token, batching still spreads fixed energy across more tokens, and context length still erodes the batch gain. Scheduling policies therefore need per-platform calibration for absolute energy, although we observe similar request-shape trends on the two evaluated platforms.

\textbf{Key Insight.} \textit{H200 lowers token energy consistently for the measured dense models, while MoE gains depend on model and request shape. Platform calibration therefore remains necessary even when qualitative batching and context trends are preserved.}

\begin{table*}[!t]
\centering
\caption{Llama-3.2-1B/H100 TTFT-oriented, prefill-dominated transition from host-bound to compute-heavy request shapes. \(T_{N=1}\) is the complete \(N=1\) inference-window time. Prompt tokens/J is computed as \((B\times C)/E_{\mathrm{NVML}}\). The TKLQT~\cite{character} sum is the sum of per-launch launch-to-start delays. Since queue intervals can overlap, the wall-clock time is not additive. Bold text marks configurations where \(T_{N=1}\) remains nearly constant as batch size increases.}
\label{tab:llama1b_tklqt_transition}
\small
\setlength{\tabcolsep}{10pt}
\renewcommand{\arraystretch}{1.25}
\begin{tabular}{@{} c c c c c c c c @{}}
\toprule
\textbf{\(B\)} & \textbf{\(C\)} & \textbf{\(N\)} &
\textbf{\(E_{\mathrm{NVML}}\) (mJ)} &
\textbf{TKLQT Avg.} &
\textbf{TKLQT Sum} &
\textbf{\(T_{N=1}\) (ms)} &
\textbf{Prompt tokens/J} \\
\midrule
1 & 512  & 1 & 5{,}988.9   & 8.05 \(\mu\)s       & 6.76 ms       & \textbf{25.18}    & 85.49 \\
2 & 512  & 1 & 6{,}810.4   & 24.38 \(\mu\)s      & 21.23 ms      & \textbf{25.41}    & 150.36 \\
4 & 512  & 1 & 9{,}295.6   & 668.23 \(\mu\)s     & 581.89 ms     & \textbf{28.73}    & 220.32 \\
8 & 512  & 1 & 33{,}905.2  & 9{,}507.70 \(\mu\)s   & 8{,}279.31 ms   & 45.27    & 120.81 \\
8 & 1024 & 1 & 71{,}897.2  & 40{,}383.57 \(\mu\)s  & 35{,}166.01 ms  & 111.31   & 113.94 \\
8 & 2048 & 1 & 191{,}210.2 & 145{,}693.76 \(\mu\)s & 131{,}561.46 ms & 325.74   & 85.69 \\
8 & 4096 & 1 & 686{,}698.1 & 540{,}615.08 \(\mu\)s & 637{,}925.79 ms & 1{,}175.80 & 47.70 \\
\bottomrule
\end{tabular}
\end{table*}

\begin{table}[!t]
\centering
\caption{Llama-3.2-1B/H100 diagnostic floor and isolated-replay estimates for host-bound and compute-heavy TTFT-oriented, prefill-dominated configurations. Floor and replay-active shares are normalized by measured NVML request energy and are reported separately because the diagnostic estimates are not strictly additive.}
\label{tab:llama1b_attribution}
\footnotesize
\setlength{\tabcolsep}{1.5pt}
\renewcommand{\arraystretch}{1.2}

\begin{tabular}{@{} p{2.4cm} c c c c @{}}
\toprule
\textbf{Request Regime} & \textbf{TKLQT} & \textbf{\(E_{\mathrm{NVML}}\)} & \textbf{Floor} & \textbf{Replay-active} \\
\textit{(Configuration)} & \textbf{Avg ($\mu$s)} & \textbf{(mJ)} & \textbf{share} & \textbf{share} \\
\midrule
\makecell[l]{\textbf{Host-bound}\\ \(B{=}1,C{=}512,\)\\ \(N{=}1\)} & 8.05 & 5,988.9 & 61.1\% & 13.6\% \\
\addlinespace[0.3em]
\makecell[l]{\textbf{Compute-heavy}\\ \(B{=}8,C{=}4096,\)\\ \(N{=}1\)} & 540,615.08  & 686,698.1 & 27.6\% & 60.4\% \\
\bottomrule
\end{tabular}
\end{table}

\subsection{TTFT and Long-Output Requests Occupy Distinct Operating Points}
\label{sec:5.3}
\begin{figure}[t]
    \centering
    \includegraphics[width=\linewidth, height=6.3cm]{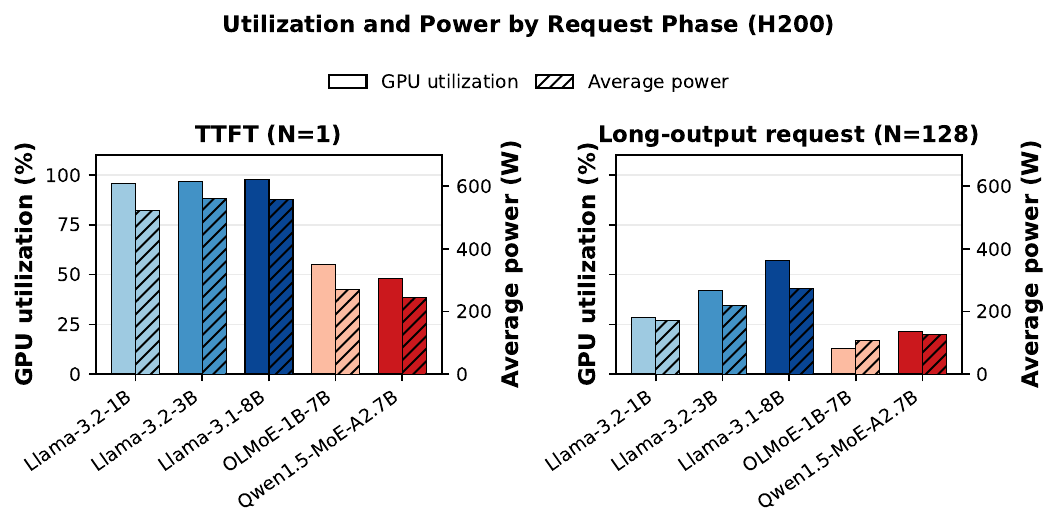}
    \caption{TTFT-oriented (\(N=1\)) and long-output
(\(N=128\)) inference operating points on H200 at
\(B=4\), \(C=2048\).}
    \label{fig:prefill_vs_decode}
\end{figure}

Fig.~\ref{fig:prefill_vs_decode} compares TTFT-oriented (\(N=1\)) and long-output (\(N=128\)) inference windows at \(B=4,C=2048\) on H200. The \(N=1\) window is prefill-dominated but also includes generation setup and first-token production, while the \(N=128\) window contains the same one-time work followed by additional decode steps. Therefore, this comparison characterizes two complete-request operating points.

\subsubsection{TTFT-oriented requests show high GPU utilization; long-output requests are lower-power but repeated}
At \(B{=}4\), \(C{=}2048\) on H200, the Llama-3.1-8B TTFT-oriented configuration (\(N{=}1\)) reaches 97.9\% GPU active-time utilization, with sampled diagnostic power of 559.1\,W, reflecting dense prompt processing over the 2{,}048-token context together with first-token production. The corresponding complete long-output request (\(N{=}128\)) has sampled diagnostic power of 273.0\,W and 56.9\% utilization, a 51.2\% difference in the smoothed power readings. Llama-3.2-1B shows the same pattern: the \(N{=}1\) configuration reaches 95.7\% utilization with sampled diagnostic power of 522.4\,W, while the \(N{=}128\) request has sampled diagnostic power of 170.2\,W and 28.6\% utilization, a 67.4\% difference. This behavior is consistent with prefill exposing substantial prompt-token parallelism, whereas the duration of a long-output request contains repeated autoregressive steps with lower GPU active-time utilization. Because NVML power telemetry on Hopper is averaged over one-second intervals, these power values characterize smoothed operating-point differences and are not used to compute request energy.

\subsubsection{Additional generation dominates long-output request energy}
At (\(B{=}16\), \(C{=}2048\)) on H200, measured Llama-3.2-1B batched inference-window energy increases from 319\,J at \(N{=}1\) to 3{,}158\,J at \(N{=}512\). The additional 2{,}839\,J relative to the TTFT-oriented configuration accounts for 89.9\% of the complete \(N{=}512\) batched inference-window energy. For Llama-3.1-8B, batched inference-window energy increases from 1{,}028\,J to 11{,}455\,J; the additional 10{,}427\,J accounts for 91.0\% of the \(N{=}512\) total. Thus, lower sampled diagnostic power during long-output requests does not imply lower total energy: repeated autoregressive generation dominates their complete batched-window energy. These differences compare complete measured request windows and do not treat the \(N{=}1\) measurement as isolated prefill energy.

\subsubsection{MoE TTFT-oriented windows show lower GPU utilization than dense-model windows} OLMoE-1B-7B and Qwen1.5-MoE-A2.7B reach 54.9\% and 47.9\% GPU active-time utilization in the \(N{=}1\) TTFT-oriented window, compared with 95.7-97.9\% for the measured Llama-1B and Llama-8B configurations. Their sampled diagnostic power, 270\,W for OLMoE-1B-7B and 244\,W for Qwen1.5-MoE-A2.7B, is lower than the 559\,W measured for Llama-3.1-8B. Lower power does not, by itself, imply better energy efficiency. The utilization results are consistent with lower, more irregular GPU activity due to expert routing and per-expert computation~\cite{taxbreak_todo}. For mixed dense/MoE serving, this means TTFT-oriented power headroom and long-output energy efficiency can move differently across model families.

\textbf{Key Insight.} \textit{TTFT-oriented and long-output request windows occupy different operating points. The former are prefill-dominated, while repeated generation with lower GPU active-time utilization determines much of the energy of long-output requests.}

\subsection{The Energy Dynamics of Free Batching}
\label{sec:5.4}
Table~\ref{tab:llama1b_tklqt_transition} shows the bounded free-batching region using complete \(N=1\) request windows. At \(C{=}512\), increasing \(B\) from 1 to 4 nearly quadruples the prompt work while \(T_{N=1}\) remains nearly constant at 25.18-28.73\,ms. Prompt efficiency therefore rises from 85.49 to 220.32 prompt tokens/J.

The regime is bounded. At \(B{=}8\), increasing \(C\) from 512 to 4096 raises \(T_{N=1}\) from 45.27 to 1{,}175.80\,ms and \(E_{\mathrm{NVML}}\) from 33{,}905 to 686{,}698\,mJ, while prompt efficiency falls from 120.81 to 47.70 prompt tokens/J. The sum of per-launch launch-to-start delays (TKLQT~\cite{character}) grows from 8{,}279 to 637{,}926\,ms over the same span. Since individual queue intervals overlap, this sum is not elapsed overhead; its growth is consistent with a deeper device queue as execution becomes compute-heavy.

Table~\ref{tab:llama1b_attribution} illustrates the endpoints using separate floor and replay-active shares: 61.1\% and 13.6\%, respectively, for the host-bound endpoint, versus 27.6\% and 60.4\% for the compute-heavy endpoint. The comparison shows a shift from an under-filled, floor-dominant window toward compute-heavy execution.

\textbf{Key Insight.} \textit{Free batching is real but bounded. \(T_{N=1}\) shows when extra batch work is nearly latency-free, while prompt tokens/J shows whether it is energy-efficient. The replay estimates show when the energy budget shifts from floor energy toward active kernel work.}

\begin{table*}[!t]
\centering
\caption{Request-shape policy summary for energy-aware serving.}
\label{tab:scheduler_decision_compact}
\small
\setlength{\tabcolsep}{14pt}
\renewcommand{\arraystretch}{1.35}

\newcolumntype{L}{>{\raggedright\arraybackslash}X}

\begin{tabularx}{\textwidth}{@{} >{\bfseries}l L L @{}}
\toprule
\textbf{Regime} & \textbf{Observed Energy Behavior} & \textbf{Strategy for Scheduler Actions} \\
\midrule
Short-context dense decode & 
Fixed energy dominates; batching strongly lowers token energy. & 
Use moderate batching when latency allows; tune the target by model scale, architecture, context length, and SLO rather than assuming a universal batch size. \\

Long-context dense decode & 
KV-cache traffic raises step energy; batch gain erodes with context length \(C\). & 
Avoid waiting too long for large batches; optimize latency and request energy jointly. \\

MoE decode at low concurrency & 
Low-concurrency MoE behavior is consistent with routing and fragmented dispatch overheads. & 
Batch MoE requests more aggressively than dense requests at the same context length. \\

Long-output decode & 
Token energy first improves as fixed energy is spread across more tokens, then can rise when step energy dominates. & 
Track both request energy and token energy; do not optimize tokens/J alone. \\

TTFT-oriented vs. long-output requests & 
TTFT-oriented windows are prefill-dominated; repeated generation dominates long-output request energy. & 
Track both TTFT-oriented and long-output request energy within the fixed-plus-marginal model. \\
\bottomrule
\end{tabularx}
\end{table*}

\subsection{Scheduler Implications and Energy-Aware Serving Design}
\label{sec:5.5}
The preceding results translate into three scheduler-facing rules: calibrate by model and runtime, choose batch size as a function of request shape, and track request-level cost rather than relying solely on the sampled power or average token-energy metric.

\subsubsection{Model and runtime effects require calibration}
Model size alone is not a reliable energy predictor. Across Llama/H200 at (\(B{=}1\), \(C{=}512\), \(N{=}10\)), Llama-3.2-3B has 3\(\times\) the parameters of Llama-3.2-1B but costs only 1.68\(\times\) more token energy (4{,}093 vs. 2{,}431\,mJ/token), while Llama-3.1-8B has 8\(\times\) the parameters but costs only 2.47\(\times\) more. The same sub-linear trend remains at higher load: at \(B{=}16\), \(C{=}512\), \(N{=}512\), the 8B/1B token-energy ratio is 3.18\(\times\). Runtime also shifts the energy surface. For Llama-3.2-3B on H100, FlashAttention-2 reduces token energy relative to eager by 13.4\% at (\(B{=}1\), \(C{=}512\)), 31.6\% at (\(B{=}1\), \(C{=}2048\)), and 51.9\% at (\(B{=}16\), \(C{=}2048\)). The increasing benefit at larger context and batch is consistent with fused attention reducing intermediate-memory traffic and improving reuse in the GPU memory hierarchy, rather than changing the model's arithmetic work~\cite{dao2022flashattention}. Schedulers, therefore, need measured model/runtime profiles rather than parameter-count heuristics.

\subsubsection{Batch targets should be shape-aware}
Moderate batching often captures much of the available token-energy gain, but the knee is model- and context-dependent. For Llama-3.2-1B/H200 at \(C{=}512\), \(N{=}10\), moving from \(B{=}1\) to \(B{=}4\) captures 79\% of the \(B{=}16\) gain. For OLMoE-1B-7B at the same shape, \(B{=}4\) captures 81\%. The additional value of batching is consistent with routing and dispatch, leaving more fixed energy to spread across generated tokens. Architecture changes the crossover point as well: OLMoE-1B-7B becomes cheaper per output token than Llama-3.1-8B at (\(B{=}4\), \(C{=}2048\), \(N{=}10\)), with 7{,}298 vs. 8{,}235\,mJ/token, and the gap widens at \(B{=}16\).

Table~\ref{tab:batch_gain_context} summarizes the broader trend. Dense models lose most of their batching gain as \(C\) grows, while the larger gains for MoE models are consistent with expert-dispatch costs leaving more fixed energy to spread across generated tokens. Thus, the batch target should be tuned by model family, context length, output length, and latency SLO.

\subsubsection{Request-level metrics are necessary}
The fixed component is large enough that sampled power and token-normalized metrics can be misleading if reported alone. For rows with floor attribution, the median floor share is 51.6\% of measured request energy; in low-load short-decode regimes (\(B{=}1\), \(C{=}512\), \(N{=}10\)), the median rises to 79.1\%. At high-load long-output shapes such as (\(B{=}16\), \(C{=}2048\), \(N{=}512\)), it falls to 31.0\%. Smoothed diagnostic power also compresses important differences: the most token-efficient measured decode setting, Llama-3.2-1B/H200 at (\(B{=}16\), \(C{=}512\), \(N{=}128\)), consumes 189.6\,mJ/token and has sampled power of 181\,W, while the least token-efficient setting, Llama-3.1-8B/H100 at (\(B{=}1\), \(C{=}8192\), \(N{=}10\)), consumes 84{,}784\,mJ/token and has sampled power of 516\,W. A 447\(\times\) token-energy spread appears as only a 2.8\(\times\) spread in the smoothed power readings.

Table~\ref{tab:scheduler_decision_compact} summarizes the resulting policy implications: batching is most valuable when fixed energy dominates, and less valuable when long-context KV-cache traffic or very long output generation makes step energy dominate.

\textbf{Key Insight.} \textit{The same generated-token count can impose very different energy costs depending on request shape. Schedulers should therefore expose request energy, token energy, and performance per watt together, and treat batch size as an energy-control knob whose benefit depends on context length, output length, and model family.}

\begin{table*}[!t]
\centering
\caption{Utility-aware capped-reasoning results at \(B{=}1\). Energy is mean per request and utility per joule is an aggregate ratio. DS denotes DeepSeek-R1-Distill.}
\label{tab:utility_reasoning_qwen}
\footnotesize
\setlength{\tabcolsep}{2pt}
\begin{tabular*}{\textwidth}{@{\extracolsep{\fill}}lllrrrrrr@{}}
\toprule
\textbf{Benchmark} & \textbf{Model} & \textbf{Cap} & \textbf{H100 Acc.} & \textbf{H100 Energy (J)} & \textbf{H100 Utility/J} & \textbf{H200 Acc.} & \textbf{H200 Energy (J)} & \textbf{H200 Utility/J} \\
\midrule
MATH-500 & Qwen3-8B & 64 & 26.0 & 316.8 & 0.000821 & 25.4 & 245.8 & 0.00103 \\
 & & 256 & 39.0 & 1{,}024.7 & 0.000381 & 39.0 & 796.3 & 0.000490 \\
 & & 1024 & 68.8 & 2{,}351.2 & 0.000293 & 68.2 & 1{,}845.8 & 0.000369 \\
 & DS-Qwen-7B & 64 & 9.4 & 640.7 & 0.000147 & 9.2 & 515.2 & 0.000179 \\
 & & 256 & 23.2 & 1{,}211.4 & 0.000192 & 21.4 & 977.9 & 0.000219 \\
 & & 1024 & 38.0 & 3{,}148.8 & 0.000121 & 36.0 & 2{,}567.4 & 0.000140 \\
 & DS-Llama-8B & 64 & 5.6 & 760.5 & 0.000074 & 5.6 & 551.4 & 0.000102 \\
 & & 256 & 6.2 & 1{,}365.0 & 0.000045 & 7.0 & 1{,}002.6 & 0.000070 \\
 & & 1024 & 13.4 & 2{,}805.1 & 0.000048 & 12.2 & 2{,}063.7 & 0.000059 \\
\midrule
ARC-Challenge & Qwen3-8B & 64 & 90.0 & 272.6 & 0.003302 & 89.7 & 200.7 & 0.004467 \\
 & & 256 & 93.7 & 950.0 & 0.000986 & 93.3 & 696.1 & 0.001341 \\
 & & 1024 & 94.0 & 1{,}169.9 & 0.000804 & 94.0 & 856.3 & 0.001098 \\
 & DS-Qwen-7B & 64 & 63.3 & 491.1 & 0.001290 & 63.3 & 394.0 & 0.001607 \\
 & & 256 & 73.0 & 1{,}023.0 & 0.000714 & 75.0 & 833.6 & 0.000900 \\
 & & 1024 & 78.7 & 2{,}035.9 & 0.000386 & 80.0 & 1{,}695.6 & 0.000472 \\
 & DS-Llama-8B & 64 & 28.0 & 759.2 & 0.000369 & 29.7 & 546.3 & 0.000543 \\
 & & 256 & 57.3 & 1{,}355.7 & 0.000423 & 61.3 & 985.5 & 0.000622 \\
 & & 1024 & 78.3 & 2{,}276.8 & 0.000344 & 79.0 & 1{,}683.4 & 0.000469 \\
\bottomrule
\end{tabular*}
\end{table*}

\section{Utility-Aware Reasoning Evaluation}
\label{app:utility}
\label{sec:5.6}

The tokenomics economic model~\cite{nvidia_tokenomics_guide} separates the utility delivered by generated tokens from the infrastructure cost of producing them, with the latter depending on request shape. This motivates a complementary question: \textit{Do additional reasoning tokens deliver utility proportional to their energy cost?} Prior work couples task accuracy with inference energy through accuracy- or intelligence-per-joule metrics~\cite{saadfalcon2025intelligence} and characterizes the energy cost of test-time reasoning~\cite{jin2025energy,oviedo2026energy}. Building on this framing, we operationalize utility as final-answer correctness and use utility per joule to compare controlled reasoning-token budgets.

We evaluate Qwen3-8B, DeepSeek-R1-Distill-Qwen-7B, and DeepSeek-R1-Distill-Llama-8B on 500 MATH-500 questions and 300 ARC-Challenge questions using greedy decoding. All three models are evaluated on both H100 and H200 GPUs. The prompt and decoding configuration are fixed within each model, and only the maximum reasoning budget is varied across 64-, 256-, and 1024-token caps. The final-answer token allowance remains fixed. Because a model can terminate reasoning before reaching its cap, the measured energy reflects the actual generated sequence rather than the nominal cap alone. Each configuration uses \(B{=}1\) and is averaged over three runs. For every response, we record final answer correctness, reasoning and answer tokens, truncation status, and request energy from the cumulative NVML energy counter.

For each recorded response \(i\), let \(U_i\in\{0,1\}\) denote final-answer correctness and \(E_i\) its measured request energy. We define utility per joule as 
\begin{equation} 
\label{eq:utility_per_joule}
    \eta_U
    =
    \frac{\sum_{i=1}^{R} U_i}
         {\sum_{i=1}^{R} E_i},
\end{equation}
where \(R\) is the number of recorded responses. Thus, \(\eta_U\) measures aggregate correct responses per joule. Accuracy is computed over distinct benchmark questions, while the energy reported in Table~\ref{tab:utility_reasoning_qwen} is mean energy per request. Unlike tokens/J, utility/J accounts for whether the generated output successfully completes the task. Hence, a system can generate tokens efficiently while producing little additional benchmark utility.

Table~\ref{tab:utility_reasoning_qwen} reveals distinct model-workload reasoning regimes. For Qwen3-8B on MATH-500, increasing the reasoning cap from 64 to 1024 raises accuracy from 26.0\% to 68.8\% on H100 and from 25.4\% to 68.2\% on H200. However, request energy increases by 7.42\(\times\) and 7.51\(\times\), respectively, reducing utility/J by approximately 2.8\(\times\) on both platforms. The diminishing return is more pronounced when short-budget accuracy is already high. For Qwen3-8B on ARC-Challenge, the same increase in cap improves accuracy by only 4.0 \% points on H100 and 4.3 \% points on H200, while request energy increases by approximately 4.3\(\times\). Consequently, utility/J decreases by approximately 4.1\(\times\) on both GPUs.

An intermediate reasoning budget is beneficial when its accuracy gain is sufficient to offset the additional energy. On MATH-500, DeepSeek-R1-Distill-Qwen-7B achieves its highest utility/J at the 256-token cap on both platforms. Similarly, DeepSeek-R1-Distill-Llama-8B achieves its highest ARC-Challenge utility/J at 256 tokens, where the additional reasoning substantially improves accuracy relative to the 64-token cap. In contrast, DeepSeek-R1-Distill-Llama-8B remains poorly matched to MATH-500: although its accuracy increases at 1024 tokens, the gain is insufficient to compensate for the additional request energy, and the 64-token cap remains utility-optimal. The utility-maximizing caps are therefore 64, 256, and 64 tokens for Qwen3-8B, DeepSeek-R1-Distill-Qwen-7B, and DeepSeek-R1-Distill-Llama-8B, respectively, on MATH-500, and 64, 64, and 256 tokens on ARC-Challenge. Notably, the 1024-token cap yields the highest accuracy for every model-benchmark pair, but does not maximize utility/J for any of them.

The platform comparison separates hardware efficiency from reasoning-budget efficiency. Across the evaluated configurations, H200 reduces mean request energy by 16.7-28.0~\% relative to H100
and improves utility/J at every operating point, while the observed accuracy differs by at most 4.0~\% points. Nevertheless, the utility-optimal cap is identical on H100 and H200 for every model-benchmark pair. Thus, greater hardware efficiency lowers the absolute energy cost of reasoning but does not eliminate the diminishing utility of additional tokens or change the preferred reasoning depth in these experiments.

\textbf{Key Insight.} \textit{Reasoning budgets should be selected jointly by model, workload, and service objective rather than by accuracy or token count alone. Short caps are appropriate when accuracy has already saturated or when the model is poorly matched to the task, whereas moderate caps are justified when their marginal accuracy gain outweighs their additional request energy. Larger reasoning budgets should be reserved for settings in which accuracy requirements take priority over energy-normalized utility. Energy-aware serving systems should therefore expose accuracy, request energy, and utility/J together when selecting models and reasoning budgets.}

\section{Summary and Conclusion}
\label{sec:conclusion}

This work presents an empirical characterization and energy model for GPU energy consumption during LLM inference.
We evaluate this LLM inference energy model on NVIDIA H100 and H200 GPUs across dense and mixture-of-experts (MoE) models, reporting both request energy and token energy as functions of request-shape parameters including:  model type \(M\), phase \(P\), batch size \(B\), context length \(C\), and output length \(N\). 
We characterize LLM inference energy as a request-shape problem rather than a static per-token cost. Using cumulative NVML energy measurements on H100 and H200 GPUs, we show that request energy and token energy can move in opposite directions: longer outputs and larger batches can reduce token energy by spreading fixed energy across more tokens, while increasing total request energy as the window extends. We also show that prefill has a bounded free-batching regime, where extra prompt work improves prompt tokens/J before larger contexts move execution into a compute-heavy regime. The scheduler-facing implication is that energy-aware serving should expose request energy, token energy, and performance per watt alongside latency, throughput, batch size, context length, and output length.
Finally, we operationalize the \textit{utility/J} metric. 
Reasoning tokens can lead to waste (saturating accuracy at certain token counts). Hence, models and the token counts can be configured for both high efficiency and low energy wastage.



\bstctlcite{IEEEexample:BSTcontrol}
\bibliographystyle{IEEEtran}
\bibliography{reference}

@IEEEtranBSTCTL{IEEEexample:BSTcontrol,
  CTLuse_article_number     = "yes",
  CTLuse_paper              = "yes",
  CTLuse_forced_etal        = "no",
  CTLmax_names_forced_etal  = "10",
  CTLnames_show_etal        = "1",
  CTLuse_alt_spacing        = "yes",
  CTLalt_stretch_factor     = "4",
  CTLdash_repeated_names    = "yes",
  CTLname_format_string     = "{f.~}{vv~}{ll}{, jj}",
  CTLname_latex_cmd         = "",
  CTLname_url_prefix        = "[Online]. Available:"
}

@misc{openai_pricing,
  author       = {{OpenAI}},
  title        = {{API Pricing}},
  howpublished = {\url{https://openai.com/api/pricing/}},
  year         = {2026},
  note         = {Accessed: 2026-05-08}
}

@inproceedings{distserve,
author = {Yinmin Zhong and Shengyu Liu and Junda Chen and Jianbo Hu and Yibo Zhu and Xuanzhe Liu and Xin Jin and Hao Zhang},
title = {{DistServe}: Disaggregating Prefill and Decoding for Goodput-optimized Large Language Model Serving},
booktitle = {18th USENIX Symposium on Operating Systems Design and Implementation (OSDI 24)},
year = {2024},
isbn = {978-1-939133-40-3},
address = {Santa Clara, CA},
pages = {193--210},
url = {https://www.usenix.org/conference/osdi24/presentation/zhong-yinmin},
publisher = {USENIX Association},
month = jul
}

@inproceedings{sarathi,
author = {Amey Agrawal and Nitin Kedia and Ashish Panwar and Jayashree Mohan and Nipun Kwatra and Bhargav Gulavani and Alexey Tumanov and Ramachandran Ramjee},
title = {Taming {Throughput-Latency} Tradeoff in {LLM} Inference with {Sarathi-Serve}},
booktitle = {18th USENIX Symposium on Operating Systems Design and Implementation (OSDI 24)},
year = {2024},
isbn = {978-1-939133-40-3},
address = {Santa Clara, CA},
pages = {117--134},
url = {https://www.usenix.org/conference/osdi24/presentation/agrawal},
publisher = {USENIX Association},
month = jul
}

@INPROCEEDINGS{dynamollm,
  author={Stojkovic, Jovan and Zhang, Chaojie and Goiri, Inigo and Torrellas, Josep and Choukse, Esha},
  booktitle={2025 IEEE International Symposium on High Performance Computer Architecture (HPCA)}, 
  title={DynamoLLM: Designing LLM Inference Clusters for Performance and Energy Efficiency}, 
  year={2025},
  volume={},
  number={},
  pages={1348-1362},
  doi={10.1109/HPCA61900.2025.00102}}

@INPROCEEDINGS{samsi2023words,
  author={Samsi, Siddharth and Zhao, Dan and McDonald, Joseph and Li, Baolin and Michaleas, Adam and Jones, Michael and Bergeron, William and Kepner, Jeremy and Tiwari, Devesh and Gadepally, Vijay},
  booktitle={2023 IEEE High Performance Extreme Computing Conference (HPEC)}, 
  title={From Words to Watts: Benchmarking the Energy Costs of Large Language Model Inference}, 
  year={2023},
  volume={},
  number={},
  pages={1-9},
  doi={10.1109/HPEC58863.2023.10363447}}

@INPROCEEDINGS{chen2024empirical,
  author={Chen, Zhen and Lin, Weiran and Xie, Xinyu and Hu, Yaodong and Li, Chao and Tong, Qiaojuan and Wu, Yinjun and Li, Shuangshou},
  booktitle={2024 IEEE International Conference on Big Data (BigData)}, 
  title={An Empirical Study on the Power Consumption of LLMs with Different GPU Platforms}, 
  year={2024},
  volume={},
  number={},
  pages={8640-8642},
  doi={10.1109/BigData62323.2024.10825662}}

@article{tokenpowerbench, title={TokenPowerBench: Benchmarking the Power Consumption of LLM Inference}, volume={40}, url={https://ojs.aaai.org/index.php/AAAI/article/view/40535}, DOI={10.1609/aaai.v40i38.40535}, abstractNote={Large language model (LLM) services now answer billions of queries per day, and industry reports show that inference, not training, accounts for more than 90% of total power consumption. However, existing benchmarks focus on either training/fine-tuning or performance of inference and provide little support for power consumption measurement and analysis of inference. We introduce TokenPowerBench, the first lightweight and extensible benchmark designed for LLM-inference power consumption studies. The benchmark combines a declarative configuration interface covering model choice, prompt set, and inference engine, a measurement layer that captures GPU-, node-, and system-level power without specialized power meters, and a phase-aligned metrics pipeline that attributes energy to the prefill and decode stages of every request. These elements make it straightforward to explore the power consumed by an LLM inference run; furthermore, by varying batch size, context length, parallelism strategy and quantization, users can quickly assess how each setting affects joules per token and other energy-efficiency metrics. We evaluate TokenPowerBench on four of the most widely used model series (Llama, Falcon, Qwen, and Mistral). Our experiments cover from 1 billion parameters up to the frontier-scale Llama3-405B model. Furthermore, we release TokenPowerBench as open source to help users to measure power consumption, forecast operating expenses, and meet sustainability targets when deploying LLM services.}, number={38}, journal={Proceedings of the AAAI Conference on Artificial Intelligence}, author={Niu, Chenxu and Zhang, Wei and Li, Jie and Zhao, Yongjian and Wang, Tongyang and Wang, Xi and Chen, Yong}, year={2026}, month={Mar.}, pages={32582–32590} }

@INPROCEEDINGS{character,
  author={Vellaisamy, Prabhu and Labonte, Thomas and Chakraborty, Sourav and Turner, Matt and Sury, Samantika and Shen, John Paul},
  booktitle={2025 IEEE International Symposium on Performance Analysis of Systems and Software (ISPASS)}, 
  title={Characterizing and Optimizing LLM Inference Workloads on CPU-GPU Coupled Architectures}, 
  year={2025},
  volume={},
  number={},
  pages={49-61},
  doi={10.1109/ISPASS64960.2025.00015}}

@misc{taxbreak_todo,
      title={TaxBreak: Unmasking the Hidden Costs of LLM Inference Through Overhead Decomposition}, 
      author={Prabhu Vellaisamy and Shreesh Tripathi and Vignesh Natarajan and Surya Santhan Thenarasu and Shawn Blanton and John P. Shen},
      year={2026},
      eprint={2603.12465},
      archivePrefix={arXiv},
      primaryClass={cs.DC},
      url={https://arxiv.org/abs/2603.12465}, 
}

@misc{executionidle2026,
      title={The Energy Cost of Execution-Idle in GPU Clusters}, 
      author={Yiran Lei and Jared Fernandez and Vasilis Kypriotis and Dimitrios Skarlatos and Emma Strubell and Justine Sherry and Daniel Vosler},
      year={2026},
      eprint={2604.04745},
      archivePrefix={arXiv},
      primaryClass={cs.DC},
      url={https://arxiv.org/abs/2604.04745}, 
}

@misc{mcdaniel2026finegrainedpowerenergyattribution,
      title={Fine-Grained Power and Energy Attribution on AMD GPU/APU-Based Exascale Nodes}, 
      author={Adam McDaniel and Michael Jantz and Ashesh Sharma and Steve Abbott and Steven Martin and Shreyas Khandekar and Brandon Neth and Bruno Villasenor Alvarez and Aditya Kashi and Wael Elwasif and Oscar Hernandez},
      year={2026},
      eprint={2604.06056},
      archivePrefix={arXiv},
      primaryClass={cs.DC},
      url={https://arxiv.org/abs/2604.06056}, 
}

@INPROCEEDINGS{mlperf,
  author={Reddi, Vijay Janapa and Cheng, Christine and Kanter, David and Mattson, Peter and Schmuelling, Guenther and Wu, Carole-Jean and Anderson, Brian and Breughe, Maximilien and Charlebois, Mark and Chou, William and Chukka, Ramesh and Coleman, Cody and Davis, Sam and Deng, Pan and Diamos, Greg and Duke, Jared and Fick, Dave and Gardner, J. Scott and Hubara, Itay and Idgunji, Sachin and Jablin, Thomas B. and Jiao, Jeff and John, Tom St. and Kanwar, Pankaj and Lee, David and Liao, Jeffery and Lokhmotov, Anton and Massa, Francisco and Meng, Peng and Micikevicius, Paulius and Osborne, Colin and Pekhimenko, Gennady and Rajan, Arun Tejusve Raghunath and Sequeira, Dilip and Sirasao, Ashish and Sun, Fei and Tang, Hanlin and Thomson, Michael and Wei, Frank and Wu, Ephrem and Xu, Lingjie and Yamada, Koichi and Yu, Bing and Yuan, George and Zhong, Aaron and Zhang, Peizhao and Zhou, Yuchen},
  booktitle={2020 ACM/IEEE 47th Annual International Symposium on Computer Architecture (ISCA)}, 
  title={MLPerf Inference Benchmark}, 
  year={2020},
  volume={},
  number={},
  pages={446-459},
  doi={10.1109/ISCA45697.2020.00045}}

@inproceedings{dao2022flashattention,
 author = {Dao, Tri},
 booktitle = {International Conference on Learning Representations},
 editor = {B. Kim and Y. Yue and S. Chaudhuri and K. Fragkiadaki and M. Khan and Y. Sun},
 pages = {35549--35562},
 title = {FlashAttention-2: Faster Attention with Better Parallelism and Work Partitioning},
 url = {https://proceedings.iclr.cc/paper_files/paper/2024/file/98ed250b203d1ac6b24bbcf263e3d4a7-Paper-Conference.pdf},
 volume = {2024},
 year = {2024}
}

@inproceedings{kwon2023pagedattention,
author = {Kwon, Woosuk and Li, Zhuohan and Zhuang, Siyuan and Sheng, Ying and Zheng, Lianmin and Yu, Cody Hao and Gonzalez, Joseph and Zhang, Hao and Stoica, Ion},
title = {Efficient Memory Management for Large Language Model Serving with PagedAttention},
year = {2023},
isbn = {9798400702297},
publisher = {Association for Computing Machinery},
address = {New York, NY, USA},
url = {https://doi.org/10.1145/3600006.3613165},
doi = {10.1145/3600006.3613165},
booktitle = {Proceedings of the 29th Symposium on Operating Systems Principles},
pages = {611–626},
numpages = {16},
location = {Koblenz, Germany},
series = {SOSP '23}
}

@misc{grattafiori2024llama,
      title={The Llama 3 Herd of Models}, 
      author={Aaron Grattafiori and Abhimanyu Dubey and Abhinav Jauhri and Abhinav Pandey and Abhishek Kadian and Ahmad Al-Dahle and Aiesha Letman and Akhil Mathur and Alan Schelten and Alex Vaughan and Amy Yang and Angela Fan and Anirudh Goyal and Anthony Hartshorn and Aobo Yang and Archi Mitra and Archie Sravankumar and Artem Korenev and Arthur Hinsvark and Arun Rao and Aston Zhang and Aurelien Rodriguez and Austen Gregerson and Ava Spataru and Baptiste Roziere and Bethany Biron and Binh Tang and Bobbie Chern and Charlotte Caucheteux and Chaya Nayak and Chloe Bi and Chris Marra and Chris McConnell and Christian Keller and Christophe Touret and Chunyang Wu and Corinne Wong and Cristian Canton Ferrer and Cyrus Nikolaidis and Damien Allonsius and Daniel Song and Danielle Pintz and Danny Livshits and Danny Wyatt and David Esiobu and Dhruv Choudhary and Dhruv Mahajan and Diego Garcia-Olano and Diego Perino and Dieuwke Hupkes and Egor Lakomkin and Ehab AlBadawy and Elina Lobanova and Emily Dinan and Eric Michael Smith and Filip Radenovic and Francisco Guzmán and Frank Zhang and Gabriel Synnaeve and Gabrielle Lee and Georgia Lewis Anderson and Govind Thattai and Graeme Nail and Gregoire Mialon and Guan Pang and Guillem Cucurell and Hailey Nguyen and Hannah Korevaar and Hu Xu and Hugo Touvron and Iliyan Zarov and Imanol Arrieta Ibarra and Isabel Kloumann and Ishan Misra and Ivan Evtimov and Jack Zhang and Jade Copet and Jaewon Lee and Jan Geffert and Jana Vranes and Jason Park and Jay Mahadeokar and Jeet Shah and Jelmer van der Linde and Jennifer Billock and Jenny Hong and Jenya Lee and Jeremy Fu and Jianfeng Chi and Jianyu Huang and Jiawen Liu and Jie Wang and Jiecao Yu and Joanna Bitton and Joe Spisak and Jongsoo Park and Joseph Rocca and Joshua Johnstun and Joshua Saxe and Junteng Jia and Kalyan Vasuden Alwala and Karthik Prasad and Kartikeya Upasani and Kate Plawiak and Ke Li and Kenneth Heafield and Kevin Stone and Khalid El-Arini and Krithika Iyer and Kshitiz Malik and Kuenley Chiu and Kunal Bhalla and Kushal Lakhotia and Lauren Rantala-Yeary and Laurens van der Maaten and Lawrence Chen and Liang Tan and Liz Jenkins and Louis Martin and Lovish Madaan and Lubo Malo and Lukas Blecher and Lukas Landzaat and Luke de Oliveira and Madeline Muzzi and Mahesh Pasupuleti and Mannat Singh and Manohar Paluri and Marcin Kardas and Maria Tsimpoukelli and Mathew Oldham and Mathieu Rita and Maya Pavlova and Melanie Kambadur and Mike Lewis and Min Si and Mitesh Kumar Singh and Mona Hassan and Naman Goyal and Narjes Torabi and Nikolay Bashlykov and Nikolay Bogoychev and Niladri Chatterji and Ning Zhang and Olivier Duchenne and Onur Çelebi and Patrick Alrassy and Pengchuan Zhang and Pengwei Li and Petar Vasic and Peter Weng and Prajjwal Bhargava and Pratik Dubal and Praveen Krishnan and Punit Singh Koura and Puxin Xu and Qing He and Qingxiao Dong and Ragavan Srinivasan and Raj Ganapathy and Ramon Calderer and Ricardo Silveira Cabral and Robert Stojnic and Roberta Raileanu and Rohan Maheswari and Rohit Girdhar and Rohit Patel and Romain Sauvestre and Ronnie Polidoro and Roshan Sumbaly and Ross Taylor and Ruan Silva and Rui Hou and Rui Wang and Saghar Hosseini and Sahana Chennabasappa and Sanjay Singh and Sean Bell and Seohyun Sonia Kim and Sergey Edunov and Shaoliang Nie and Sharan Narang and Sharath Raparthy and Sheng Shen and Shengye Wan and Shruti Bhosale and Shun Zhang and Simon Vandenhende and Soumya Batra and Spencer Whitman and Sten Sootla and Stephane Collot and Suchin Gururangan and Sydney Borodinsky and Tamar Herman and Tara Fowler and Tarek Sheasha and Thomas Georgiou and Thomas Scialom and Tobias Speckbacher and Todor Mihaylov and Tong Xiao and Ujjwal Karn and Vedanuj Goswami and Vibhor Gupta and Vignesh Ramanathan and Viktor Kerkez and Vincent Gonguet and Virginie Do and Vish Vogeti and Vítor Albiero and Vladan Petrovic and Weiwei Chu and Wenhan Xiong and Wenyin Fu and Whitney Meers and Xavier Martinet and Xiaodong Wang and Xiaofang Wang and Xiaoqing Ellen Tan and Xide Xia and Xinfeng Xie and Xuchao Jia and Xuewei Wang and Yaelle Goldschlag and Yashesh Gaur and Yasmine Babaei and Yi Wen and Yiwen Song and Yuchen Zhang and Yue Li and Yuning Mao and Zacharie Delpierre Coudert and Zheng Yan and Zhengxing Chen and Zoe Papakipos and Aaditya Singh and Aayushi Srivastava and Abha Jain and Adam Kelsey and Adam Shajnfeld and Adithya Gangidi and Adolfo Victoria and Ahuva Goldstand and Ajay Menon and Ajay Sharma and Alex Boesenberg and Alexei Baevski and Allie Feinstein and Amanda Kallet and Amit Sangani and Amos Teo and Anam Yunus and Andrei Lupu and Andres Alvarado and Andrew Caples and Andrew Gu and Andrew Ho and Andrew Poulton and Andrew Ryan and Ankit Ramchandani and Annie Dong and Annie Franco and Anuj Goyal and Aparajita Saraf and Arkabandhu Chowdhury and Ashley Gabriel and Ashwin Bharambe and Assaf Eisenman and Azadeh Yazdan and Beau James and Ben Maurer and Benjamin Leonhardi and Bernie Huang and Beth Loyd and Beto De Paola and Bhargavi Paranjape and Bing Liu and Bo Wu and Boyu Ni and Braden Hancock and Bram Wasti and Brandon Spence and Brani Stojkovic and Brian Gamido and Britt Montalvo and Carl Parker and Carly Burton and Catalina Mejia and Ce Liu and Changhan Wang and Changkyu Kim and Chao Zhou and Chester Hu and Ching-Hsiang Chu and Chris Cai and Chris Tindal and Christoph Feichtenhofer and Cynthia Gao and Damon Civin and Dana Beaty and Daniel Kreymer and Daniel Li and David Adkins and David Xu and Davide Testuggine and Delia David and Devi Parikh and Diana Liskovich and Didem Foss and Dingkang Wang and Duc Le and Dustin Holland and Edward Dowling and Eissa Jamil and Elaine Montgomery and Eleonora Presani and Emily Hahn and Emily Wood and Eric-Tuan Le and Erik Brinkman and Esteban Arcaute and Evan Dunbar and Evan Smothers and Fei Sun and Felix Kreuk and Feng Tian and Filippos Kokkinos and Firat Ozgenel and Francesco Caggioni and Frank Kanayet and Frank Seide and Gabriela Medina Florez and Gabriella Schwarz and Gada Badeer and Georgia Swee and Gil Halpern and Grant Herman and Grigory Sizov and Guangyi and Zhang and Guna Lakshminarayanan and Hakan Inan and Hamid Shojanazeri and Han Zou and Hannah Wang and Hanwen Zha and Haroun Habeeb and Harrison Rudolph and Helen Suk and Henry Aspegren and Hunter Goldman and Hongyuan Zhan and Ibrahim Damlaj and Igor Molybog and Igor Tufanov and Ilias Leontiadis and Irina-Elena Veliche and Itai Gat and Jake Weissman and James Geboski and James Kohli and Janice Lam and Japhet Asher and Jean-Baptiste Gaya and Jeff Marcus and Jeff Tang and Jennifer Chan and Jenny Zhen and Jeremy Reizenstein and Jeremy Teboul and Jessica Zhong and Jian Jin and Jingyi Yang and Joe Cummings and Jon Carvill and Jon Shepard and Jonathan McPhie and Jonathan Torres and Josh Ginsburg and Junjie Wang and Kai Wu and Kam Hou U and Karan Saxena and Kartikay Khandelwal and Katayoun Zand and Kathy Matosich and Kaushik Veeraraghavan and Kelly Michelena and Keqian Li and Kiran Jagadeesh and Kun Huang and Kunal Chawla and Kyle Huang and Lailin Chen and Lakshya Garg and Lavender A and Leandro Silva and Lee Bell and Lei Zhang and Liangpeng Guo and Licheng Yu and Liron Moshkovich and Luca Wehrstedt and Madian Khabsa and Manav Avalani and Manish Bhatt and Martynas Mankus and Matan Hasson and Matthew Lennie and Matthias Reso and Maxim Groshev and Maxim Naumov and Maya Lathi and Meghan Keneally and Miao Liu and Michael L. Seltzer and Michal Valko and Michelle Restrepo and Mihir Patel and Mik Vyatskov and Mikayel Samvelyan and Mike Clark and Mike Macey and Mike Wang and Miquel Jubert Hermoso and Mo Metanat and Mohammad Rastegari and Munish Bansal and Nandhini Santhanam and Natascha Parks and Natasha White and Navyata Bawa and Nayan Singhal and Nick Egebo and Nicolas Usunier and Nikhil Mehta and Nikolay Pavlovich Laptev and Ning Dong and Norman Cheng and Oleg Chernoguz and Olivia Hart and Omkar Salpekar and Ozlem Kalinli and Parkin Kent and Parth Parekh and Paul Saab and Pavan Balaji and Pedro Rittner and Philip Bontrager and Pierre Roux and Piotr Dollar and Polina Zvyagina and Prashant Ratanchandani and Pritish Yuvraj and Qian Liang and Rachad Alao and Rachel Rodriguez and Rafi Ayub and Raghotham Murthy and Raghu Nayani and Rahul Mitra and Rangaprabhu Parthasarathy and Raymond Li and Rebekkah Hogan and Robin Battey and Rocky Wang and Russ Howes and Ruty Rinott and Sachin Mehta and Sachin Siby and Sai Jayesh Bondu and Samyak Datta and Sara Chugh and Sara Hunt and Sargun Dhillon and Sasha Sidorov and Satadru Pan and Saurabh Mahajan and Saurabh Verma and Seiji Yamamoto and Sharadh Ramaswamy and Shaun Lindsay and Shaun Lindsay and Sheng Feng and Shenghao Lin and Shengxin Cindy Zha and Shishir Patil and Shiva Shankar and Shuqiang Zhang and Shuqiang Zhang and Sinong Wang and Sneha Agarwal and Soji Sajuyigbe and Soumith Chintala and Stephanie Max and Stephen Chen and Steve Kehoe and Steve Satterfield and Sudarshan Govindaprasad and Sumit Gupta and Summer Deng and Sungmin Cho and Sunny Virk and Suraj Subramanian and Sy Choudhury and Sydney Goldman and Tal Remez and Tamar Glaser and Tamara Best and Thilo Koehler and Thomas Robinson and Tianhe Li and Tianjun Zhang and Tim Matthews and Timothy Chou and Tzook Shaked and Varun Vontimitta and Victoria Ajayi and Victoria Montanez and Vijai Mohan and Vinay Satish Kumar and Vishal Mangla and Vlad Ionescu and Vlad Poenaru and Vlad Tiberiu Mihailescu and Vladimir Ivanov and Wei Li and Wenchen Wang and Wenwen Jiang and Wes Bouaziz and Will Constable and Xiaocheng Tang and Xiaojian Wu and Xiaolan Wang and Xilun Wu and Xinbo Gao and Yaniv Kleinman and Yanjun Chen and Ye Hu and Ye Jia and Ye Qi and Yenda Li and Yilin Zhang and Ying Zhang and Yossi Adi and Youngjin Nam and Yu and Wang and Yu Zhao and Yuchen Hao and Yundi Qian and Yunlu Li and Yuzi He and Zach Rait and Zachary DeVito and Zef Rosnbrick and Zhaoduo Wen and Zhenyu Yang and Zhiwei Zhao and Zhiyu Ma},
      year={2024},
      eprint={2407.21783},
      archivePrefix={arXiv},
      primaryClass={cs.AI},
      url={https://arxiv.org/abs/2407.21783}, 
}

@inproceedings{fernandez2025energy,
    title = "Energy Considerations of Large Language Model Inference and Efficiency Optimizations",
    author = "Fernandez, Jared  and
      Na, Clara  and
      Tiwari, Vashisth  and
      Bisk, Yonatan  and
      Luccioni, Sasha  and
      Strubell, Emma",
    editor = "Che, Wanxiang  and
      Nabende, Joyce  and
      Shutova, Ekaterina  and
      Pilehvar, Mohammad Taher",
    booktitle = "Proceedings of the 63rd Annual Meeting of the Association for Computational Linguistics (Volume 1: Long Papers)",
    month = jul,
    year = "2025",
    address = "Vienna, Austria",
    publisher = "Association for Computational Linguistics",
    url = "https://aclanthology.org/2025.acl-long.1563/",
    doi = "10.18653/v1/2025.acl-long.1563",
    pages = "32556--32569",
    ISBN = "979-8-89176-251-0"
}

@inproceedings{wilhelm2025energytoken,
  author    = {Wilhelm, Patrick and Wittkopp, Thorsten and Kao, Odej},
  title     = {{Beyond Test-Time Compute Strategies: Advocating Energy-per-Token in LLM Inference}},
  booktitle = {Proceedings of the 5th Workshop on Machine Learning and Systems},
  series    = {EuroMLSys '25},
  year      = {2025},
  pages     = {208--215},
  publisher = {Association for Computing Machinery},
  doi       = {10.1145/3721146.3721953},
  url       = {https://doi.org/10.1145/3721146.3721953}
}

@inproceedings{sweetspots2026,
   title={SweetSpot: An Analytical Model for Predicting Energy Efficiency of LLM Inference},
   url={http://dx.doi.org/10.1145/3777884.3797011},
   DOI={10.1145/3777884.3797011},
   booktitle={Proceedings of the 17th ACM/SPEC International Conference on Performance Engineering},
   publisher={ACM},
   author={Pizzini Cavagna, Hiari and Proia, Andrea and Madella, Giacomo and Esposito, Giovanni Battista and Antici, Francesco and Cesarini, Daniele and Kiziltan, Zeynep and Bartolini, Andrea},
   year={2026},
   month=May, pages={83–95} }

@article{oviedo2026energy,
   title={Energy use of AI inference, efficiency pathways, and test-time scaling},
   ISSN={2542-4351},
   url={http://dx.doi.org/10.1016/j.joule.2026.102430},
   DOI={10.1016/j.joule.2026.102430},
   journal={Joule},
   publisher={Elsevier BV},
   author={Oviedo, Felipe and Kazhamiaka, Fiodar and Choukse, Esha and Kim, Allen and Luers, Amy and Nakagawa, Melanie and Bianchini, Ricardo and Lavista Ferres, Juan M.},
   year={2026},
   month=Apr, pages={102430} }

@manual{nvml2026,
  title        = {{NVML API Reference Guide}},
  author       = {{NVIDIA Corporation}},
  organization = {{NVIDIA Corporation}},
  year         = {2026},
  note         = {NVIDIA GPU Deployment and Management Documentation},
  url          = {https://docs.nvidia.com/deploy/nvml-api/index.html}
}

@misc{nvidia-ml-py,
  title        = {{nvidia-ml-py}},
  author       = {{nvidia-ml-py contributors}},
  howpublished = {PyPI package. \url{https://pypi.org/project/nvidia-ml-py/}},
  year         = {2026}
}

@misc{torch_profiler2026,
  author       = {{PyTorch Contributors}},
  title        = {{PyTorch Profiler Tool}},
  howpublished = {Meta AI, Linux Foundation. \url{https://pytorch.org/tutorials/recipes/recipes/profiler_recipe.html}},
  year         = {2026}
}

@article{paszke2019pytorch,
  title={Pytorch: An imperative style, high-performance deep learning library},
  author={Paszke, Adam and Gross, Sam and Massa, Francisco and Lerer, Adam and Bradbury, James and Chanan, Gregory and Killeen, Trevor and Lin, Zeming and Gimelshein, Natalia and Antiga, Luca and others},
  journal={Advances in neural information processing systems},
  volume={32},
  year={2019}
}

@inproceedings{insights_llm_energy,
    title = "Towards Sustainable {NLP}: Insights from Benchmarking Inference Energy in Large Language Models",
    author = "Poddar, Soham  and
      Koley, Paramita  and
      Misra, Janardan  and
      Ganguly, Niloy  and
      Ghosh, Saptarshi",
    editor = "Chiruzzo, Luis  and
      Ritter, Alan  and
      Wang, Lu",
    booktitle = "Proceedings of the 2025 Conference of the Nations of the Americas Chapter of the Association for Computational Linguistics: Human Language Technologies (Volume 1: Long Papers)",
    month = apr,
    year = "2025",
    address = "Albuquerque, New Mexico",
    publisher = "Association for Computational Linguistics",
    url = "https://aclanthology.org/2025.naacl-long.632/",
    doi = "10.18653/v1/2025.naacl-long.632",
    pages = "12688--12704",
    ISBN = "979-8-89176-189-6"
}

@misc{energy_to_token,
      title={Position: LLM Inference Should Be Evaluated as Energy-to-Token Production}, 
      author={Xiang Liu and Shimiao Yuan and Zhenheng Tang and Peijie Dong and Kaiyong Zhao and Qiang Wang and Bo Li and Xiaowen Chu},
      year={2026},
      eprint={2605.11733},
      archivePrefix={arXiv},
      primaryClass={cs.CE},
      url={https://arxiv.org/abs/2605.11733}, 
}

@misc{liu2025greenllmsloawaredynamicfrequency,
      title={GreenLLM: SLO-Aware Dynamic Frequency Scaling for Energy-Efficient LLM Serving}, 
      author={Qunyou Liu and Darong Huang and Marina Zapater and David Atienza},
      year={2025},
      eprint={2508.16449},
      archivePrefix={arXiv},
      primaryClass={cs.PF},
      url={https://arxiv.org/abs/2508.16449}, 
}

@inproceedings{wilkins2024offline,
  title     = {Offline Energy-Optimal {LLM} Serving: Workload-Based Energy Models for {LLM} Inference on Heterogeneous Systems},
  author    = {Wilkins, Grant and Keshav, Srinivasan and Mortier, Richard},
  booktitle = {Proceedings of the 3rd Workshop on Sustainable Computer Systems (HotCarbon)},
  year      = {2024},
  eprint    = {2407.04014},
  doi       = {10.1145/3727200.3727217},
  archivePrefix = {arXiv}
}

@misc{kakolyris2025sloawaregpufrequencyscaling,
      title={SLO-aware GPU Frequency Scaling for Energy Efficient LLM Inference Serving}, 
      author={Andreas Kosmas Kakolyris and Dimosthenis Masouros and Petros Vavaroutsos and Sotirios Xydis and Dimitrios Soudris},
      year={2025},
      eprint={2408.05235},
      archivePrefix={arXiv},
      primaryClass={cs.DC},
      url={https://arxiv.org/abs/2408.05235}, 
}

@misc{ifath2026characterizingperformanceenergytradeoffslarge,
      title={Characterizing Performance-Energy Trade-offs of Large Language Models in Multi-Request Workflows}, 
      author={Md. Monzurul Amin Ifath and Israat Haque},
      year={2026},
      eprint={2604.09611},
      archivePrefix={arXiv},
      primaryClass={cs.DC},
      url={https://arxiv.org/abs/2604.09611}, 
}

@article{tran2026wattchmen,
  title={Wattchmen: Watching the Wattchers--High Fidelity, Flexible GPU Energy Modeling},
  author={Tran, Brandon and Maiterth, Matthias and Shin, Woong and Sinclair, Matthew D and Venkataraman, Shivaram},
  journal={arXiv preprint arXiv:2603.26435},
  year={2026}
}

@misc{nvidiah100specs,
  author       = {{NVIDIA}},
  title        = {{NVIDIA H100 Tensor Core GPU}},
  howpublished = {Product specifications},
  year         = {2026},
  url          = {https://www.nvidia.com/en-us/data-center/h100/},
  note         = {Accessed July 2026}
}

@misc{nvidiah200specs,
  author       = {{NVIDIA}},
  title        = {{NVIDIA H200 Tensor Core GPU}},
  howpublished = {Product specifications},
  year         = {2026},
  url          = {https://www.nvidia.com/en-us/data-center/h200/},
  note         = {Accessed July 2026}
}

@article{ujeniyah200power2026,
  author        = {Ujeniya, Aditya and Eitzinger, Jan and Hager, Georg and Wellein, Gerhard},
  title         = {{Architectural Trade-offs in the Energy-Efficient Era: A Comparative Study of Power-Capping NVIDIA H100 and H200}},
  year          = {2026},
  eprint        = {2604.11391},
  archivePrefix = {arXiv},
  primaryClass  = {cs.DC},
  url           = {https://arxiv.org/abs/2604.11391}
}

@article{mayrpoweraware2026,
  author        = {Mayr, M. and Wind, S. and Schr{\"o}der, L. and Hager, G. and K{\"o}stler, H. and Wellein, G.},
  title         = {{AI Application Benchmarking: Power-Aware Performance Analysis for Vision and Language Models}},
  year          = {2026},
  eprint        = {2603.16164},
  archivePrefix = {arXiv},
  primaryClass  = {cs.DC},
  url           = {https://arxiv.org/abs/2603.16164}
}

@misc{nvidia_tokenomics_guide,
  author       = {{NVIDIA}},
  title        = {{Tokenomics Guide: Case Studies and More}},
  howpublished = {Online documentation},
  year         = {2026},
  url          = {https://www.nvidia.com/en-us/solutions/ai/tokenomics-guide/},
  note         = {Accessed: 2026-08-25}
}

@article{saadfalcon2025intelligence,
  author  = {Jon Saad-Falcon and Avanika Narayan and
             Hakki Orhun Akengin and J. Wes Griffin and
             Herumb Shandilya and Adrian Gamarra Lafuente and
             Medhya Goel and Rebecca Joseph and Shlok Natarajan and
             Etash Kumar Guha and Shang Zhu and Ben Athiwaratkun and
             John Hennessy and Azalia Mirhoseini and Christopher R{\'e}},
  title   = {Intelligence per Watt: Measuring Intelligence
             Efficiency of Local AI},
  journal = {arXiv preprint arXiv:2511.07885},
  year    = {2025},
  doi     = {10.48550/arXiv.2511.07885}
}

@article{jin2025energy,
  author  = {Yunho Jin and Gu-Yeon Wei and David Brooks},
  title   = {The Energy Cost of Reasoning: Analyzing Energy Usage
             in LLMs with Test-Time Compute},
  journal = {arXiv preprint arXiv:2505.14733},
  year    = {2025},
  doi     = {10.48550/arXiv.2505.14733}
}

@IEEEtranBSTCTL{BSTcontrol,
  CTLuse_forced_etal       = "yes",
  CTLmax_names_forced_etal = "6",
  CTLnames_show_etal       = "1"
}

\end{document}